\documentclass[useAMS, usenatbib]{mn2e}
\usepackage{amssymb,amsmath}
\usepackage{graphicx}
\usepackage{xcolor}

\title[BLR-less AGNs]
{Optically-Selected BLR-less Active Galactic Nuclei from the SDSS Stripe82 
Database I: The Sample}
\author[Zhang X.-G.]
       {Xue-Guang Zhang$^{1,2}$\\
       $^1$Purple Mountain Observatory, Chinese Academy of Sciences,
             2 Beijing XiLu, NanJing, JiangSu, 210008, P. R. China \\
       $^2$Chinese Center for Antarctic Astronomy, NanJing,
             JiangSu, 210008, P. R. China}

\date{}
\pagerange{\pageref{firstpage}--\pageref{lastpage}} \pubyear{}
\def\LaTeX{L\kern-.36em\raise.3ex\hbox{a}\kern-.15em
    T\kern-.1667em\lower.7ex\hbox{E}\kern-.125emX}
\label{firstpage}

\date{}
\def\LaTeX{L\kern-.36em\raise.3ex\hbox{a}\kern-.15em
    T\kern-.1667em\lower.7ex\hbox{E}\kern-.125emX}

\begin{document}
\pagerange{\pageref{firstpage}--\pageref{lastpage}} \pubyear{}
\maketitle
\label{firstpage}

\begin{abstract}
     This is the first paper in a dedicated series to study 
the properties of the optically selected BLR-less AGNs (Active Galactic 
Nuclei with no-hidden central broad emission line regions). We carried 
out a systematic search for the BLR-less AGNs through the Sloan Digital 
Sky Survey Legacy Survey (SDSS Stripe82 Database). Based on the spectral 
decomposition results for all the 136676 spectroscopic objects (galaxies 
and QSOs) with redshift less than 0.35 covered by the SDSS Stripe82 
region, our spectroscopic sample for the BLR-less AGNs includes 
22693 pure narrow line objects without broad emission lines but with 
apparent AGN continuum emission $R_{AGN}>0.3$ and apparent stellar 
lights $R_{ssp}>0.3$. Then, using the properties of the photometry magnitude 
RMS ($RMS$) and the Pearson's coefficients ($R_{1, 2}$) between two 
different SDSS band light curves: $RMS_k>3\times RMS_{M_k}$ and 
$R_{1, 2}>\sim0.8$, the final 281 pure narrow objects with true photometry 
variabilities are our selected reliable candidates for the BLR-less AGNs. 
The selected candidates with higher confidence levels not only have 
the expected spectral features of the BLR-less AGNs, but also show 
significant true photometry variabilities. The reported sample at least 
four-times enlarges the current sample of the BLR-less AGNs, and will 
provide more reliable information to explain the lack of the BLRs of 
AGNs in our following studies.
\end{abstract}


\section{Introduction}
    For Active Galactic Nuclei (AGN), the well-known constantly being 
revised Unified Model (\citealt{rees84, mg90, an93, up95, zw06, ss07, 
wz07, hi09, sr11, e12, mb12, mbb12}, more recent review can be 
found in \citealt{bm12}) can be applied to explain most of the different 
observed phenomena of different kinds of AGN, due to the different 
orientation angles of the central accretion disk combining with the 
different central accretion rates, the different covering factors 
of the dust torus etc.. Based on the Unified Model, one simple viewpoint 
is that type 2 AGN (no broad emission lines in observed spectra) are 
intrinsically like type 1 AGN (apparent broad lines in observed spectra), 
but their broad-line regions (BLRs) are hidden from our view by the dust 
torus and/or high density dust clouds. This simple viewpoint can be strongly 
supported by the clearly detected polarized broad emission lines and/or 
by the clearly detected broad infrared broad emission lines for some 
type 2 AGN (\citealt{mg90, ah94, kay94, yh96, hl97, vs97, km98, ya98, 
bf99a, bf99b, vs99, au00, mb00, lg00, tr01, ss02, tr03, nk04}). 

   However, the increasing number of studies, especially on the 
analysis of the X-ray band characters and the analysis on the 
polarimetric spectral properties for some type 2 objects, have shown 
that besides the type 1 AGN (including those seriously obscured objects, 
such as type 1.5, type 1.9 AGN etc.) and the type 2 AGN, there is one 
special kind of AGN, BLR-less AGN (named as true type 2 AGN or AGN 
without hidden BLRs in some references): AGN without central BLRs 
(\citealt{pb02, l03,  gz03, ha04, gp07, bc08, bn08, sr10, mbb12, 
py12}). Based on the polarimetric spectral features and/or based on 
the photometric magnitude variabilities, around 70 BLR-less AGNs have 
been reported and discussed in the literature. Besides the 
contributions to improve the Unified Model for AGN, a study the 
properties of the BLR-less AGNs should provide more information on  
the formation (or the suppression) of the BLR of AGN, and/or additional 
information about transition stages of the accretion etc. 
(\citealt{l03, nm03, wg05, sb06, gs07, ha07, eh09, pc09, c10, 
gp10, pab10, sr10, tr11, mb12, py12}).

    So far, there have been several methods proposed to explain the BLR 
absence in AGNs. Based on the spectropolarimetric sample from Tran (2001), 
\citet{nm03} have shown that the absence or presence of the hidden BLR 
in the type 2 AGNs is controlled by the central accretion, under the 
assumption that the BLRs are formed by the accretion disk instabilities 
occurring around the critical radius at which the gas pressure dominated 
accretion disk changes to being dominated by the radiation pressure 
(\citealt{ni00}). \citet{l03} have shown that BLR can not be able to 
survive, if central luminosity is much lower. \citet{eh09} have shown 
that the disk-wind scenario for BLR and torus obscuration predicts 
the disappearance of the BLR for AGNs at low luminosities. More recently, 
\citet{c10} have shown that for the low luminosity AGNs containing ADAF, 
the inner small cold disk is evaporated completely and the outer thin 
accretion disk may be seriously suppressed, which leads to the 
lack of the BLR. However, based on the study of the current sample of 
the BLR-less AGNs, the reason of the BLR absence in BLR-less AGN is 
still controversial. \citet{mb12, py12} supported the model proposed 
by \citet{nm03}, through a sample of 18 BLR-less AGNs (\citealt{mb12}) 
and a sample of 36 BLR-less AGNs (\citealt{py12}). \citet{zw06} have 
found that the BLR absence in the BLR-less AGNs is probably caused by the 
less massive black holes and the high accretion rates similar to those in 
Narrow Line Seyfert I objects, through a sample of 46 BLR-less AGNs. 
\citet{wz07} have shown that the lack of BLR could be probably caused 
by low gas-to-dust ratios, through a sample of 69 BLR-less AGNs. 
\citet{bg07} have shown that the AGN luminosity plays a major role 
in the BLR absence in the BLR-less AGNs, while for the high-luminosity 
BLR-less AGNs, the BLR absence depends not only on the AGN activity, but 
also on the torus obscuration, through a sample of 49 BLR-less AGNs. 
\citet{wz11} have shown that the BLR absence in the luminous BLR-less AGNs 
depends on the obscuration, however, the BLR absence the less luminous 
BLR-less AGNs depends on the very low Eddington ratio rather than the 
obscuration, based on a sample of 71 BLR-less AGNs. \citet{yh11} have 
shown that in order to explain the nitrogen overabundance for the 
BLR-less AGNs, there should be apparent effects from stellar evolution, 
based on a sample of 33 BLR-less AGNs.

    Due to the current small sample of the BLR-less AGNs reported in the 
literature, some contradictory statements about the properties of the 
BLR-less AGNs have been reported, which arise mainly due to the limited 
size of the current sample of the BLR-less AGNs and/or due to the 
different methods to select the candidates for the BLR-less AGNs. Thus, 
in this manuscript, we increase the current sample of the BLR-less AGNs. 
It is clear that the most direct method to select the BLR-less AGNs is 
through the polarimetric spectral properties (such as the sample in 
Tran 2001), and through the continuum variabilities of the pure narrow 
line objects (such as the sample in \citealt{ha04}). Fortunately, the 
Sloan Digital Sky Survey (SDSS, \citealt{ya00, am08, ab09, aa11}) 
Stripe82 database, covering the region with the right ascension from 
20.7h to 3.3h and with the declination from -1.26 to 1.26, provides an idea 
to collect the BLR-less AGNs through the properties of the SDSS spectra 
and the photometric variabilities. 

    The manuscript is organized as follows. In Section 2, we describe 
our procedures to collect the candidates for the BLR-less AGNs, based on 
both the SDSS spectral features and the properties of the photometry 
variabilities through the SDSS Stripe82 database. Then, our discussions, 
conclusions and our simple and basic results about the primary 
parameters of our BLR-less AGNs are shown in the Section 3. The 
cosmological parameters $H_{0}=70{\rm km\cdot s}^{-1}{\rm Mpc}^{-1}$, 
$\Omega_{\Lambda}=0.7$ and $\Omega_{m}=0.3$ have been adopted.

\section{Data Sample for BLR-less AGNs}

   In order to select the reliable candidates for the BLR-less AGNs 
through both the photometry variabilities and the observed spectral 
features, we applies the following procedures to the objects 
in the SDSS Stripe82 Database. The emission line parameters should 
be firstly determined, in order to confirm that our spectroscopic  sample 
includes the pure narrow line objects only (type 2 objects, the objects 
having both strong and weak broad lines can not be considered). Then, 
the SDSS photometric light curves of the objects in the spectroscopic sample 
should be carefully analysed to find the reliable candidates for the 
BLR-less AGNs.

\subsection{Line Parameters for the Spectroscopic Objects in the 
SDSS Stripe82 Region}

    The SDSS Stripe82 region consists of two scan regions referred to 
as the north and the south strips. Both the north and the south strips 
have been repeatedly imaged since 1998. A brief description of the 
Stripe82 data and coadd can be found in \citet{ab09}. Some detailed 
descriptions about the constructions for the corrected photometric 
light-motion curves for the objects in the Stripe82 region can be 
found in \citet{vb07}, \citet{is07}, \citet{bv08} and the references 
therein. Using the properties of the photometric light curves of the 
spectroscopic objects in the Stripe82 region, we are coming to find 
the reliable candidates for the BLR-less AGNs.  

   Because the line parameters (especially, line width and line flux) 
are necessary for us to create our spectroscopic sample for the BLR-less 
AGNs: only pure narrow line objects with high quality SDSS spectra 
are considered in our spectroscopic sample. The following two data files have 
been collected from the SDSS website 
(http://www.sdss3.org/dr8/spectro/spectro\_access.php): 
'specObj-dr8.fits' which provides the information about the redshifts, 
the classifications, the positions etc. for all the SDSS spectroscopic 
objects in the SDSS DR8, and 'galSpecIndx-dr8.fits' which provides the 
information about the line parameters for all the SDSS spectroscopic 
objects in the SDSS DR8. Then based on the information of the RA and the 
DEC for all the objects in the SDSS DR8: $300<RA<60$ and $-1.25<DEC<1.25$, 
229318 SDSS spectroscopic objects with redshift less than 0.35 covered by 
the Stripe82 region are firstly selected. Here, the restriction on  
the redshift is applied to ensure the spectra of the collected objects 
cover both H$\alpha$ and H$\beta$, in order to check whether there are 
broad lines. Based on the SDSS pipeline, there are 114788 of 229318 
objects classified as 'GALAXY', 92642 of 229318 objects classified as 
'STAR' and 21888 of 229318 objects classified as 'QSO'.  

   Then, by the selected 229318 spectroscopic objects covered in the 
Stripe82 region, it is straightforward to create our spectroscopic sample 
including the pure narrow line objects, based on the measured line 
parameters for all the objects except the stars covered by the Stripe82 
region. Due to the contributions from the stellar lights, the spectral 
decomposition should be firstly applied. In other words, before to measure 
the line parameters of the emission lines, the contributions of the 
stars to the SDSS spectrum should be subtracted. Here, the most commonly 
accepted "Simple Stellar Population" (SSP) method is applied: the stellar 
contributions in the observed SDSS spectrum can be best fitted and 
subtracted by the linear combination of the SSPs.  The SSP method 
provides the fundamental link between theory/models and observations: 
\begin{equation}
P_{O,\lambda} = (\sum_{j=1}^{39}A_J\times P_{ssp, \delta\lambda, r_{\lambda}})\otimes G(\lambda,\sigma) + P_{AGN, r_{\lambda}}
\end{equation} 
where $P_{O,\lambda}$ represents the observed spectrum, $\delta\lambda$ 
represents one small wavelength shift between the observed spectrum and 
the SSPs, $r_{\lambda}$ represents one intrinsic reddening factor, 
$P_{AGN, r_{\lambda}}$ represents the AGN component (commonly described by 
one power-law function) with the reddening factor  $r_{\lambda}$, 
$P_{ssp, \delta\lambda, r_{\lambda}}$ means the SSP components with the 
wavelength shift $\delta\lambda$ and with the reddening factor 
$r_{\lambda}$, $G(\lambda,\sigma)$ means one broadening function with 
the broadening velocity $\sigma$ (commonly, the stellar velocity 
dispersion).

\begin{figure*}
\centering\includegraphics[width = 18cm,height=12cm]{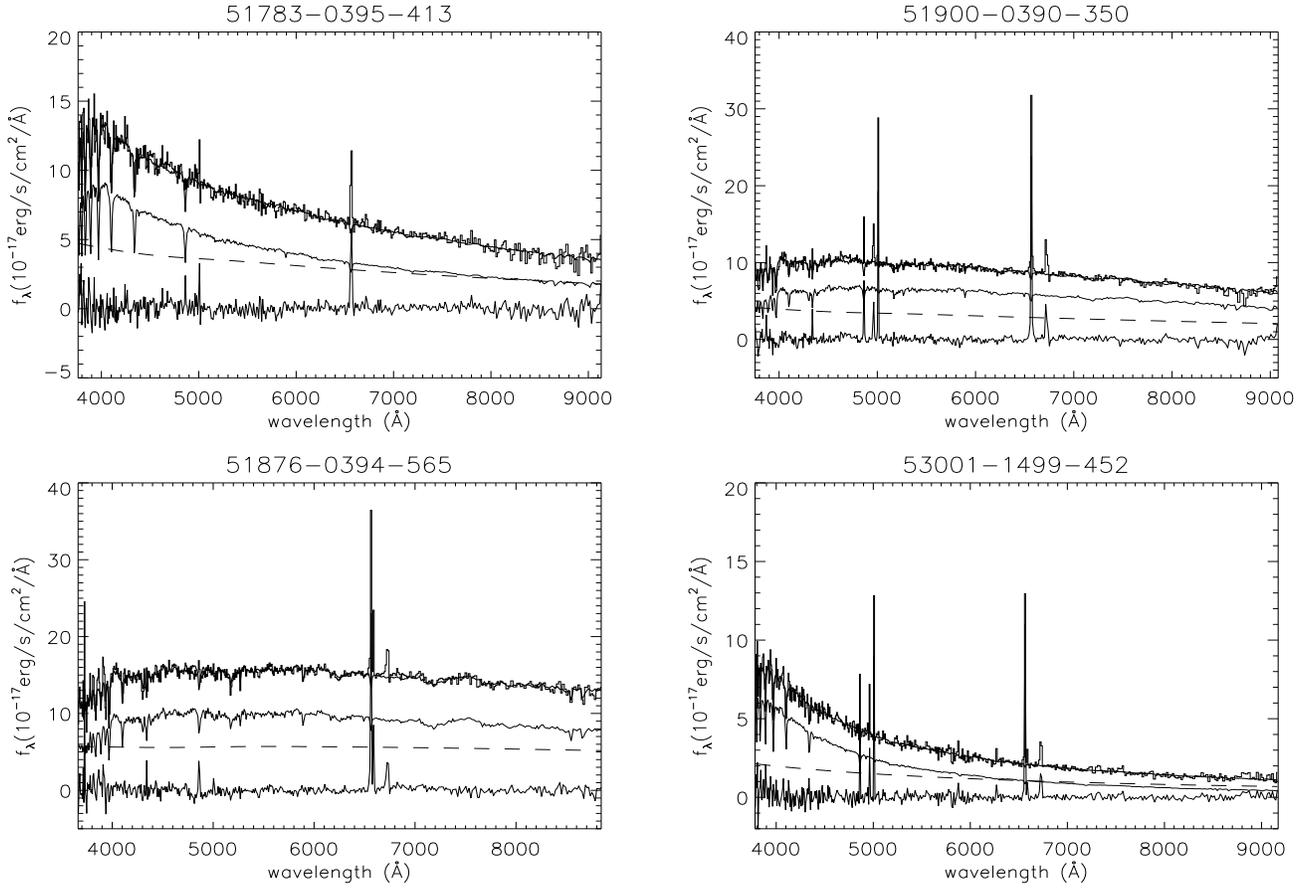}
\caption{Examples of the spectral decomposition by the SSP method. 
In each panel, the observed SDSS spectrum (thin line), the best 
fitted results (thick line), the stellar components, the AGN continuum 
(dashed line) and the line spectrum are shown (from top to bottom). 
In order to show more clearer plots, each component has been smoothed 
using 8 nearest data points. The title of each panel shows the 
information of the SDSS MJD-PLATE-FIBERID for the spectrum.
}
\label{sp_decom}
\end{figure*}

    Here, we exploit the 39 simple stellar population templates 
from the \citet{bc03}, which include the population age from 5 Myr to 
12 Gyr, with three solar metallicities (Z = 0.008, 0.05, 0.02). As the 
detailed discussion in \citet{bc03}, the 39 templates can be used to 
well-describe the characters of nearly all the galaxies in SDSS, such as 
the distributions of the stellar masses, ages, metallicities etc.. 
Then through the Levenberg-Marquardt least-squares minimization 
method applied for the SDSS spectra with the emission lines being masked, 
the stellar component and the AGN power law continuum component 
in the SDSS spectra can be clearly determined and separated. A detailed 
descriptions of the SSP method can be found in 
\citet{bc03, ka03, bc04, cg04, th04, cg05, cm05, sc06, ca07, rf09, vs10, 
gm11, ms11, pz11, bw12, cd12, cm12} etc..  There are several other methods 
than SSP  used for the spectral decomposition, such as the PCA (Principle 
Component Analysis) method (\citealt{yc04, ha05, lw05, vs06, zd08}), 
ICA (Independent Component Analysis) method (\citealt{zw06}). Each method 
has its own advantages and disadvantages. Here, the SSP method is used, 
because the SSP method results can provide the direct stellar properties, 
and moreover the SSP templates can be conveniently collected from the 
literature.  Some examples for the spectral decomposition can be found 
in Figure~\ref{sp_decom}. From the results shown in the figure, it is 
clear that both the stellar components and the power law continuum 
component can be clearly subtracted from the observed SDSS spectrum. 
The four objects shown in the figure are four reliable candidates in 
our final sample of the candidates for the BLR-less AGNs.

     Once the stellar components are subtracted out, the emission lines 
can be measured. Here, we mainly focus on the emission lines around the 
H$\alpha$ and the H$\beta$, including the optical Fe~{\sc ii} emission 
lines, the broad He~{\sc ii}$\lambda4687\AA$ line, the common/extended 
[O~{\sc iii}]$\lambda4959, 5007\AA$ doublet, the narrow/broad H$\beta$, 
the [O~{\sc i}]$\lambda6300,6363\AA$ doublet, the 
[N~{\sf ii}]$\lambda6548,6583\AA$ doublet, the 
[S~{\sc ii}]$\lambda6716, 6731\AA$ doublet and the narrow/broad H$\alpha$. 
Here, the narrow/broad H$\gamma$ and H$\delta$ lines are not considered, 
due to their weakness, which have no effects on our following results.

   It is straightforward to measure the line parameters by simple 
Gaussian functions applied to the emission lines combined with one power 
law function used for the AGN continuum, after the subtraction 
of the stellar contributions. Each narrow Gaussian function (full-width 
at half maximum less than 800${\rm km/s}$) is applied for each narrow 
emission line, except the extended components of the 
[O~{\sc iii}]$\lambda4959, 5007\AA$ doublet. For the [O~{\sc iii}] doublet, 
besides the normal narrow gaussian functions, there are two broad gaussian 
functions applied for the extended wings of the doublet as discussed 
in \citet{gh05}. For the doublets [O~{\sc iii}] and [N~{\sc ii}], the 
center wavelength ratio, flux ratio and width ratio are fixed to the 
theoretical values. For the doublets [O~{\sc i}] and [S~{\sc ii}], only 
the center wavelength ratio and the width ratio are fixed to the theoretical 
values. Two broad gaussian functions (full-width at half maximum larger 
than 600${\rm km/s}$) are applied for the broad H$\alpha$ (H$\beta$), 
i.e., one normal broad function and one much extended broad function. 
A simple broad gaussian function is sufficient to describe observed 
broad H$\alpha$ (H$\beta$) of a part of the objects. The main reason to 
describe the broad Balmer lines by two broad gaussian functions is only 
to find more better description for the broad line profile, in order to 
reject the objects with probable broad balmer lines. To further discuss 
why two broad components are much preferred for some objects are beyond 
the scope of the manuscript.  Then, one broadened Fe~{\sc ii} template 
spectra is applied for the probable Fe~{\sc ii} components within the 
wavelength range from 4100$\AA$ to 5600$\AA$ (\citealt{bg92, sp03, ko10}). 
Here, the method described in \citet{ko10} is applied to measure the 
Fe~{\sc ii} properties. Finally, using the Levenberg-Marquardt 
least-squares minimization method, the line parameters (including the 
Fe~{\sc ii}) and the characters of the AGN continuum emission can be 
determined. 

\begin{figure*}
\centering\includegraphics[width = 10cm,height=8cm]{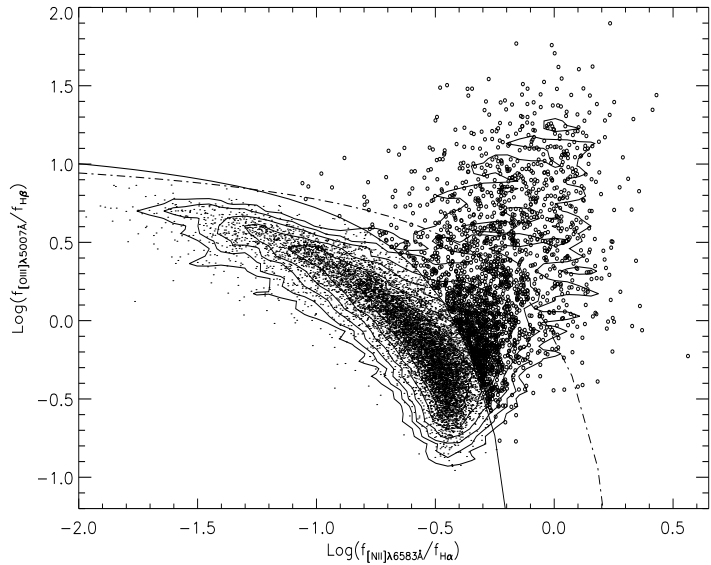}
\caption{The BPT diagram for the pure narrow line objects. The solid 
line(the dot-dashed line) represents the dividing line between the 
AGNs (Seyferts and LINERs) and the HIIs discussed in Kauffmann et 
al. (2003) and in Kewley et al. (2009). The dots and 
the circles are the classified HIIs and the AGNs. The contour is 
created by the selected pure narrow line objects.
}
\label{bpt}
\end{figure*}

    Then, based on the measured line parameters: $P\ge3\times P_{err}$ 
(where $P$ and $P_{err}$ represent the measured line parameters and the 
correspond uncertainties), among the 136676 objects (114788 objects 
classified as 'GALAXY' and 21888 objects classified as 'QSO', through the 
SDSS pipeline), 27806 pure narrow objects (67 of 21888 'QSO' objects and 
27739 of 114788 'GALAXY' objects) have apparent and strong narrow emission 
lines of [O~{\sc iii}] and [N~{\sc ii}] doublets, but no broad emission 
lines. Moreover, there are 60066 objects (394 of 21888 QSOs and 59672 of 
114788 Galaxies) having much weaker narrow emission lines and no broad lines.  
For the objects with no broad, the parameter 'subclass' listed in the SDSS 
data file 'specObj-dr8.fits' is considered, only the objects with the 
spectroscopic subclassification not including 'BROADLINE' are considered. 
Finally, the total 87872 objects (27806 pure narrow line objects, and 
60066 objects with much weaker narrow lines and no broad lines) are maken 
up of our first sample. Then, simple classification for the 87872 objects 
can be done through the well-known BPT diagram (\citealt{bpt81, vo87, ke01, 
ka03, ke06}). There are 17123 objects classified as AGNs (Seyfert galaxies 
and LINERs), 23144 objects classified as HII galaxies, and 47605 objects 
not classified due to the less information of narrow line ratios. 
Figure~\ref{bpt} shows the properties of the objects with strong narrow 
lines in the BPT diagram. If strong narrow lines are apparent, 
the commonly used line ratios of [O~{\sc iii}]/H$\beta$ and 
[N~{\sc ii}]/H$\alpha$ are used for the classification. If only one 
line ratio is available, the AGN classification can be done by 
[O~{\sc iii}]/H$\beta$ larger than 10 (or [N~{\sc ii}]/H$\alpha$ 
larger than 1).

   There are two more points we should note. On the one hand, although 
the SDSS standard pipeline output has been used to classify the objects 
into GALAXY and QSO, some objects classified as GALAXY have apparent 
broad lines, and some objects classified as QSO have apparent stellar 
components. Hence, the spectral decomposition is applied to all the 
spectroscopic objects (galaxies and QSOs) observed in the Stripe82 region. 
On the other hand, when the BPT diagram is used for the classification, 
the line ratios of [O~{\sc iii}]/H$\beta$ and [N~{\sc ii}]/H$\alpha$ are 
mainly considered. The other line ratios, such as [O~{\sc ii}]/H$\beta$ and 
[S~{\sc ii}]/H$\alpha$ are not considered, due to the weak line 
intensities and/or large effects from the extinction on the line ratios. 


\subsection{Spectroscopic Candidates for the BLR-less AGNs}

    Our spectroscopic candidates are expected to have the spectral 
features of the BLR-less AGNs, which are no broad emission lines but 
apparent power law continuum emission. Here, the spectral decomposition 
results are discussed for the pure narrow line objects with strong 
AGN continuum emissions.  

\begin{figure*}
\centering\includegraphics[width = 14cm,height=6cm]{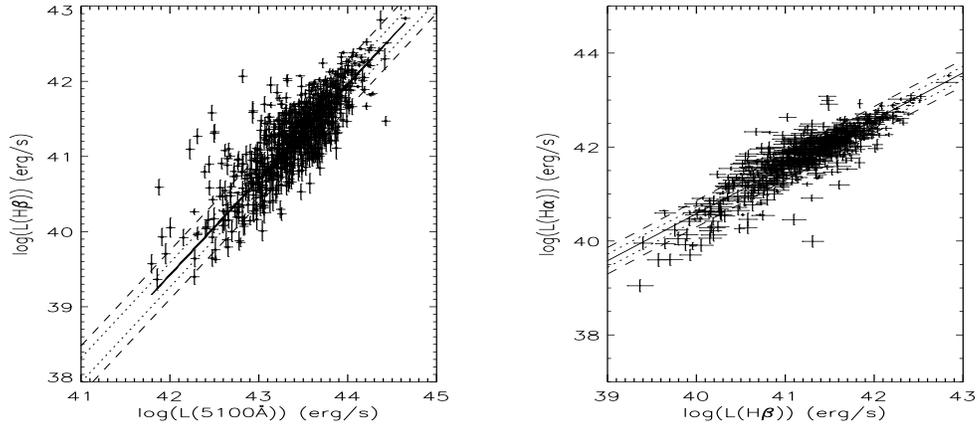}
\caption{Correlation between the AGN continuum luminosity at 5100\AA 
and the broad H$\beta$ luminosity (left panel) and the luminosity 
correlation between the broad Balmer lines (right panel), for the pure broad 
line AGNs. The solid line shows the best fitted result, the dotted line 
and the dashed line represent the 0.1dex and 0.2dex scatters for the 
corresponding best fitted result. }
\label{type1_con_broad}
\end{figure*}

   First, we check the linear correlation between the AGN 
continuum luminosity and the broad line luminosity reported by 
Greene \& Ho (2005) for pure QSOs, in order to confirm that we reliably 
measured AGN continuum luminosities. Figure~\ref{type1_con_broad} shows 
the correlation for the 815 pure type 1 AGNs of which spectra include 
both apparent broad balmer lines and apparent stellar lights in the 
Stripe82 region. The main reason for not considering the obscured broad 
line AGNs (having broad H$\alpha$ but much weak broad H$\beta$) is to 
avoid the effects of the internal extinctions. The luminosity of the 
broad H$\beta$ is the total luminosity, in spite of the broad H$\beta$ 
being described by one or two broad gaussian components. The strong linear 
correlation is present: the spearman rank correlation coefficient 0.82 
with $P_{null}\sim0$. Including the uncertainties in both coordinates, the 
linear correlation can be written as,
\begin{equation}
\log(\frac{L(H\beta)}{\rm erg/s}) = (-13.79\pm0.34) + (1.26\pm0.007)\times\log(\frac{L(5100\AA)}{{\rm erg/s}})  
\end{equation}
The result is comparable to the one in Greene \& Ho (2005), and 
indicates our spectral decomposition results are reliable. The right 
panel of the Figure~\ref{type1_con_broad} shows the strong 
luminosity correlation between the broad H$\alpha$ and the broad 
H$\beta$ for the 815 pure type 1 AGNs, which will indicate whether our 
measured line parameters are reliable. The correlation coefficient is 
about 0.87 with $P_{null}\sim0$. The best fitted result for the luminosity 
correlation is $L(H\alpha) = (3.8\pm0.1)\times L(H\beta)$, which  
indicates small internal extinction for the results in the left panel of
 Figure~\ref{type1_con_broad}. 

    Based on the spectral decomposition results, the following 
three criteria are used to select for the pure narrow line objects as 
our spectroscopic candidates for the NLR-less AGNs,
\begin{equation}
\begin{split}
&R_{AGN} > 30\% \\
&R_{ssp} > 30\% \\
&P_{line} > 3\times P_{line,err}
\end{split}
\end{equation}
where $R_{ssp} = \frac{f(ssp,5100\AA)}{f(ssp, 5100\AA)+f(AGN, 5100\AA)}$ 
and $R_{AGN} = \frac{f(AGN,5100\AA)}{f(ssp, 5100\AA)+f(AGN, 5100\AA)}$ 
mean the contributions to the continuum at 5100\AA from stellar 
light and from the AGN continuum emission, $P_{line}$ and $P_{line,err}$ 
represent the measured line parameters for at least three of the narrow 
lines (narrow H$\beta$, H$\alpha$, [O~{\sc iii}] doublet, [N~{\sc ii}] 
doublet, [S~{\sc ii}] doublet, [O~{\sc i}] doublet).  The first criterion 
ensures the existence of the AGN continuum emission, the second criterion 
confirms the stellar component, and the third criterion shows there are 
at least three normal narrow spectral emission lines. Here, the critical 
values for $R_{AGN}$ and $R_{ssp}$ are determined as follows. The value 
17\% was accepted as the standard flux uncertainty, based on the measured 
line parameters of the broad balmer lines of the 815 pure type 1 AGNs 
shown in Figure~\ref{type1_con_broad}. Therefore, the strength of the 
AGN continuum emission (stellar lights) is at least twice the uncertainties 
and indicate there are apparent and reliable AGN continuum emission 
(stellar lights), thus $R_{AGN}$ ($R_{ssp}$) should be at least larger 
than $\sim0.30$.  

\begin{figure*}
\centering\includegraphics[width = 10cm,height=8cm]{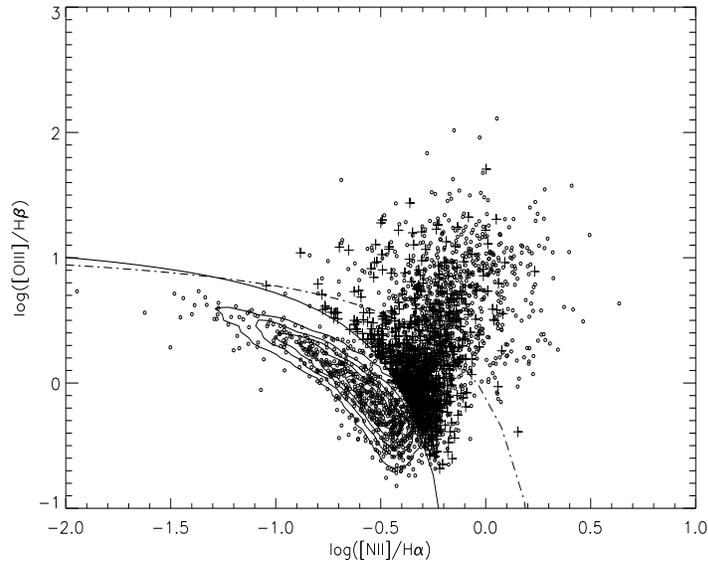}
\caption{The BPT diagram for the broad line AGNs (including the pure QSOs). 
The contour located below the dividing lines represent the positions for the 
classified HIIs from the pure narrow line objects, the plus symbol represents 
the classified AGNs from the pure narrow line objects, the circles are for 
the broad line AGNs.}
\label{broad_bpt}
\end{figure*}

    Using the three criteria above, there are 22693 narrow line 
objects selected as our spectroscopic sample of the candidates for the 
BLR-less AGNs. Among the 22693 narrow line objects, 877 objects can be 
classified as AGNs (Seyfert and LINER objects), 12577 objects can be 
classified as HII galaxies, based on the BPT line ratios of 
[O~{\sc iii}]/H$\beta$ and [N~{\sc ii}]/H$\alpha$, and 9239 objects can 
not be classified due to insufficient narrow line information. The BPT 
diagram can be used to classify the narrow emission line objects. In order 
to show that the classified HII galaxies by the BPT 
line ratios can have apparent AGN power law continuum emission, we show 
the properties of all the 1790 broad line AGNs (including the high 
luminosity QSOs) having both broad H$\alpha$ and broad H$\beta$ in the 
Stripe82 region in the BPT diagram of [O~{\sc iii}]/H$\beta$ versus 
[N~{\sc ii}]/H$\alpha$ in Figure~\ref{broad_bpt}. Because of the 
insufficient number of narrow lines, some broad line AGNs in the 
Stripe82 region are not considered in the BPT diagram. It is 
clear that even for the broad line AGN, part of the broad line objects 
lie below the decomposition line between the HII galaxies and AGNs 
(Seyfert galaxies and LINERs) (Kauffmann et al. 2003, Kewley et al. 2009). 
Similar results can also be found in Schawinski et al. (2010). 
Thus, we can reasonably accept the apparent power law continuum component 
in the spectra of part of the HII galaxies.    

   Finally, our spectroscopic sample includes 22693 pure narrow line 
objects (9239 non-classified objects and 13454 classified objects). The 
objects in the sample have apparent power law continuum emission, 
apparent narrow spectral emission lines, apparent stellar lights, but no 
broad lines. The spectral features of the objects are akin to  
the expected spectral features of the BLR-less AGNs. Besides the 
spectral features described above,  the properties of the photometric 
variabilities should be used to find the candidates with high confidence 
levels for the BLR-less AGNs, because variability is one of the fundamental 
characters of AGN.


\subsection{Variability-Selection of the BLR-less AGNs}

    It is convenient to check the properties of the photometry 
variabilities of the objects in our spectroscopic sample. The 
photometry light curves spanning over seven years in the Stripe82 
database are discussed in detail in \citet{bv08}. Through the database 
provided by Bramich et al. (2008, 
http://das.sdss.org/va/stripe\_82\_variability/SDSS\_82\_public/),
we can conveniently collect the the SDSS five band light curves (with
exponential and point spread function (PSF) magnitudes along with
uncertainties) and the necessary corresponding parameters on the light
curves (such as the rms scatter, object type, Stetson and Vidrih 
variability indices etc.) for all the objects in the SDSS Stripe82 
region. Then, based on the photometry variability properties, the 
following procedures are applied to find the reliable candidates for 
the BLR-less AGNs.

   It is commonly and well known that there are two important parameters 
to describe the photometry variabilities and to determine and select the 
true variable objects: the RMS photometry magnitude deviation ($RMS$) as 
the function of the magnitude (see the results in Bramich et al. 2008, 
Kozlowshi et al. 2010, and references therein) and the correlation 
coefficient $R_{1, 2}$ between two different band light curves (see 
more recent definition in Kozlowski et al. 2010). In this subsection, we 
check the properties of the two parameters for the pure narrow line 
objects in our spectroscopic sample, in order to find objects with true and 
reliable variabilities, which will be our final candidates for the 
BLR-less AGNs. Certainly, in order to find more reliable RMS magnitude 
deviations ($RMS$) and correlation coefficients, only the objects having 
more than 10 reliable photometry observations are considered. 

\begin{figure*}
\centering\includegraphics[width = 18cm,height=20cm]{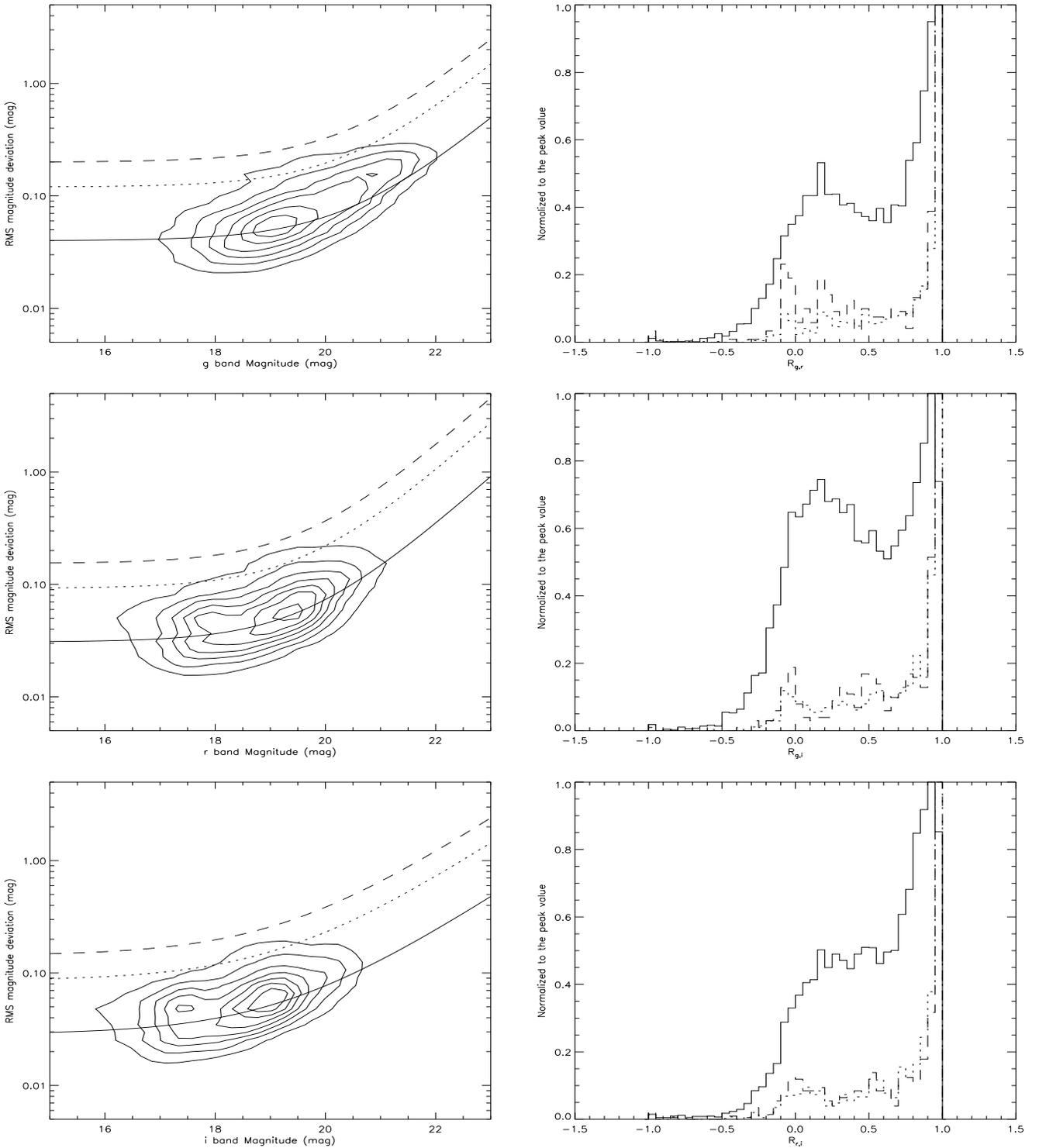}
\caption{Properties of the RMS functions (left panels) and the 
Pearson's coefficients between two different band light curves 
(right panels) for all the galaxies and QSOs in the Stripe82 region.  
In the left panel, the contour (the correlation between the RMS deviation 
and the magnitude) shows all the objects (9254 QSOs and 
51614 galaxies for the g band results, 9826 QSOs and 55795 
galaxies for the r band results, 9728 QSOs and 54695 galaxies for the 
i band results), the solid line, the dotted line and the dashed line 
shows $RMS_{k}/RMS_{M_k}=1,3,5$ ($RMS_{k}/RMS_{M_k}=1$ means the 
results shown in the Equation (4)) respectively. In the right panel, 
the solid line, the dotted line and the dashed line represent the 
properties of the Pearson's coefficients for the objects with 
$RMS_{k}/RMS_{M_k}>1$ and for the objects with $RMS_{k}/RMS_{M_k}>3$ 
and for the objects with $RMS_{k}/RMS_{M_k}>5$ respectively.   
}
\label{mag_rms}
\end{figure*}

   Before proceeding further, some simple descriptions are shown on 
the parameters of $RMS$ and $R_{1, 2}$. Here, the parameter $R_{1, 2}$ 
is the Pearson's correlation coefficient between two different band 
light curves, an expected feature of truly variable objects. And, the 
dependence of the function of $RMS$ on photometry magnitude is described 
as $RMS=A+B\times exp(C\times(Mag - 18))$ as posed in Bramich et al. 
(2008), and then determined as follows.  In the plane of photometry 
magnitude and the RMS, all the spectroscopic objects (galaxies) 
in the Stripe82 region are binned into 0.1mag. And moreover, in order 
to ignore the effects of the objects with large variabilities, we only 
consider the objects having their own standard deviations less than 
the corresponding RMS magnitude deviation for all objects in the 
corresponding bin. Then the RMS functions for the SDSS bands are 
found as
\begin{equation}
\begin{split}
RMS_g &= 0.039+0.004\times exp(0.96\times(Mag_g - 18)) \\
RMS_r &= 0.031+0.006\times exp(1.01\times(Mag_r - 18)) \\
RMS_i &= 0.028+0.012\times exp(0.74\times(Mag_i - 18))
\end{split}
\end{equation}
Due to the some significantly poorer signal-to-noise for SDSS u and 
z bands, we mainly consider and show the results on the SDSS gri bands 
in the manuscript. Figure~\ref{mag_rms} shows the corresponding results 
about the $RMS$ (the solid line the left panels), $R_{g,r}$ (correlation 
between g band and r band light curves), $R_{g,i}$ (correlation 
between g band and i band light curves) and $R_{r,i}$ (correlation 
between r band and i band light curves) for all the galaxies and QSOs 
in the Stripe82 region.

   Then, the determined $RMS$ function and the calculated $R_{1, 2}$ can 
be used as the variability indicators, and will mean larger RMS values and 
higher Pearson's coefficients, for true variabilities:
\begin{equation}
\begin{split} 
RMS_{k, 1} &> k_{RMS} \times RMS_{1,M_k} \\
RMS_{k, 2} &> k_{RMS} \times RMS_{2,M_k} \\
R_{1, 2} &> R_{critical} 
\end{split}
\end{equation},
where $RMS_{k, band}$ and $RMS_{band, M_k}$ mean the RMS magnitude   
deviation for the $kth$ object in the given SDSS band and the calculated 
RMS photometry magnitude deviation calculated by the $RMS$ function 
given the mean photometry magnitude in the given band of the object. 
Follow the procedures in Kozlowski et al. (2010), we estimate the fraction 
of the false-positives among the selected objects with true variabilities by
\begin{equation}
f_{fake} = \frac{1-R_{critical}}{1.5}\times \frac{N_{R_{1, 2}<0.5}}{N_{R_{1, 2}>R_{critical}}}  
\end{equation}
where $N_{R_{1,2}<0.5}$ ($N_{R_{1,2}>R_{critical}}$) means the number 
of the selected objects with Pearson's coefficients between two given band 
light curves smaller than 0.5 (larger than $R_{critical}$). Here, we 
accept that the weak- or un-correlated objects have the Pearson's 
coefficient smaller than 0.5. Similar to we adopt, in Kozlowski et al. (2010), 
the critical value is $R_{critical}=0.8$ for highly correlated objects. 
However, in the manuscript, we require the determined $f_{fake}$ to  
be smaller than 10\% for our following selected candidates for the 
BLR-less AGNs, which leads to $RMS_{k} > 3\times RMS_{M_k}$ for 
three SDSS bands, $R_{r,i}>R_{critical}\sim0.75$ for selecting 
candidates for the BLR-less AGNs through the r and i band results, 
$R_{g,r}>R_{critical}\sim0.85$ for selecting the candidates through the 
g and r band results and $R_{g,i}>R_{critical}\sim0.85$ for selecting the 
candidates through the g and i band results. Here, the value 
$k_{RMS}>3$ strongly guarantees the variability significance.  Then based 
on the critical values $R_{critical}$ and the calculated $RMS$ values, 
the reliable 281 candidates for the BLR-less AGNs can be selected. 

\subsection{The Final Sample}

\begin{figure*}
\centering\includegraphics[width = 18cm,height=20cm]{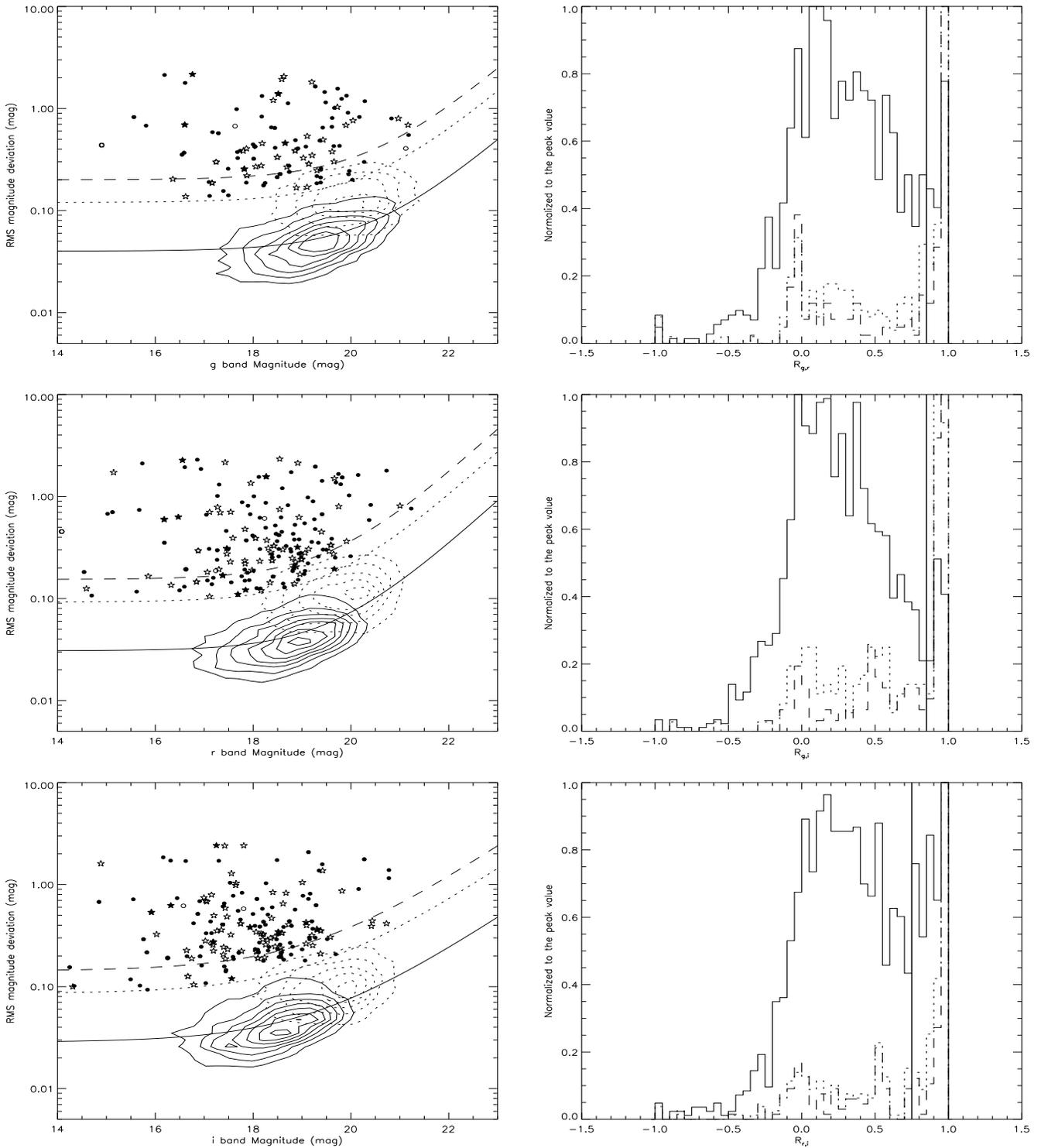}
\caption{Similar as the results in Figure~\ref{mag_rms}, but for the 
objects in our spectroscopic sample. And furthermore, in the left 
panels, the  contour in solid line shows the properties of the 
narrow line objects in our spectroscopic sample, the contour in 
dotted line shows the properties of the QSOs in Stripe82 region. And in 
the left panels, all the 281 reliable candidates are shown with open 
circles (AGNs), solid circles (HII galaxies) and triangles (non-classified 
objects). In the right panels, the thick solid lines show the positions 
for the critical value of $R_{1, 2}$. 
}
\label{mag_rms2}
\end{figure*}

   The 281 reliable candidates for the BLR-less AGNs have the following 
feature: {\bf true and apparent AGN continuum variabilities, 
apparent narrow emission lines, and no broad emission lines}. Among the 
281 candidates, there are 9 AGNs, 171 HII galaxies and 101 
non-classified objects nu the SDSS. Meanwhile, the expected number of 
false-positives among the candidates is less than 30 ($f_{fake}$ 
less than 10\%). The properties of the RMS functions for the reliable 
candidates are shown in the left panels in Figure~\ref{mag_rms2}. 
Because only two band light curves out of three three band (SDSS gri bands) 
light curves are available for some of the candidates. Therefore, in 
the figure, 133 of the 281 candidates are shown in top left panel, 
202 of the 281 candidates are shown in the middle left panel, and 209 
of 281 candidates are shown in the bottom left panel. 
Figure~\ref{mag_rms2} also shows the normal narrow line objects in the 
spectroscopic sample (contour in solid line in the figure) and all the 
spectroscopic QSOs with redshift less than 0.35 (contour in dotted 
line in the figure) in the Stripe82 region. The larger variabilities 
($RMS_{k}/RMS_{M_k}$) of the candidates than those of a large part of 
QSOs provide further evidence for the true variabilities of 
the candidates.

     Finally, based on both the spectral characters and the photometry 
variabilities, 281 reliable candidates are included in our final sample. 
The basic information of the candidates is listed in Table 1, including 
the SDSS MJD-PLATE-FIBERID information, the redshift, the position, 
the r band magnitude, the values of $R_{AGN}$, $R_{1,2}$, $k_{rms}$ and 
classification. 

\section{Discussions and Conclusions}

     Based on the photometry variabilities of the spectroscopic objects 
from the SDSS Stripe82 region, we selected 281 candidates for the 
BLR-less AGNs with high confidence levels. We discuss three points/caveats 
of our spectral analysis below.

\begin{figure*}
\centering\includegraphics[width = 10cm,height=8cm]{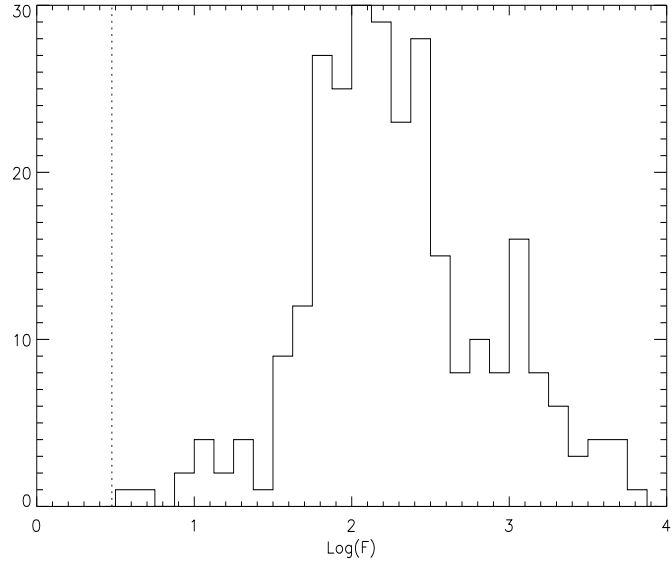}
\caption{The F-test results. The solid line is for the distribution 
for the calculated F values by the Equation (7), the vertical dotted 
line is the F-value calculated by the F distribution. 
}
\label{ftest}
\end{figure*}

    First and foremost, the determined power law AGN continuum component 
is one necessary parameter to select the candidates for the BLR-less AGNs, 
based on the definition of the BLR-less AGN: the nuclei are directly 
observed. Thus, the reliability of the AGN continuum determination based 
on the spectral decomposition, should be carefully further discussed, 
besides the results shown in the Figure~\ref{type1_con_broad}. The sum 
of the pure SSPs gives one power law continuum feature, unless the 
broadening velocity for the SSPs is unreasonably large or only the much 
younger SSPs are included. Thus, the parameter $R_{AGN}>0.3$ 
mathematically support the reliability of the power law AGN continuum 
component. Furthermore, in order to provide clearer evidence for 
the reliability of the AGN continuum determination, the procedure in 
the subsection 2.1 is re-applied without the considerations of the AGN 
continuum component $P_{AGN, r_{\lambda}}$ for the observed spectra 
of the final 281 candidates for the BLR-less AGNs in our final sample.  
Then, the F-test technique is applied to check the reliability of the 
AGN component $P_{AGN, r_{\lambda}}$, 
\begin{equation}
F = \frac{(SSE_1 - SSE_2)/(DoF_1-DoF_2)}{SSE_2/(DoF_2)}
\end{equation}
where $SSE$ represents the sum of squared residuals for one model,
the suffix '1' is for the simple model and '2' is for the slightly
complicated model, $DoF$ represents the degree of freedom for one model.
Then, the calculated value $F$ should be compared to the F-value estimated
by the F-distribution with the numerator degrees of freedom of
$DoF_1 - DoF_2$ and the denominator degrees of freedom of $DoF_2$. It is
clear that the F-value by the F-distribution with $p=0.05$ is $\sim3$ 
based on the numerator and denominator degrees of freedom. The 
calculated $F$ values through the equation above are much larger than 3, 
as the shown results in the Figure~\ref{ftest}. Thus, the mathematical 
F-test results support the AGN component $P_{AGN, r_{\lambda}}$.  

\begin{figure*}
\centering\includegraphics[width = 18cm,height=10cm]{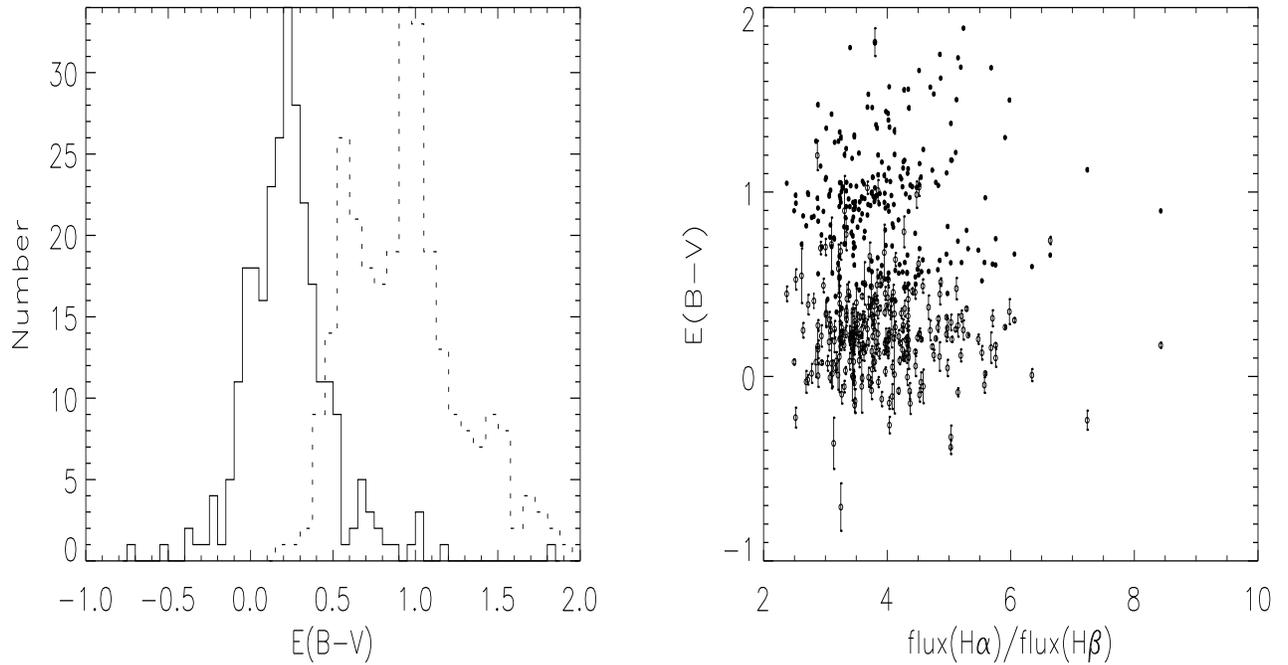}
\caption{The properties of the dust extinctions with and without the 
considerations of the AGN continuum component $P_{AGN, r_{\lambda}}$ for 
the 281 reliable candidates for the BLR-less AGNs. In the left panel, 
the solid line is for the results with the consideration of 
$P_{AGN, r_{\lambda}}$, the dotted line is for the results without the 
consideration of $P_{AGN, r_{\lambda}}$. In the right panel, it is shown 
the correlation between the flux ratio of the narrow H$\alpha$ to the 
narrow H$\beta$ and the parameter E(B-V). In the right panel, open 
circles are for the results with the consideration of 
$P_{AGN, r_{\lambda}}$, solid circles are for the results without 
the consideration of $P_{AGN, r_{\lambda}}$.
}
\label{ebv}
\end{figure*}

    Second, we simply check the properties of the the intrinsic dust 
extinction included in the procedure in the subsection 2.1. A detailed  
discussions of the correlation between the dust extinctions and the other 
stellar parameters are beyond the scope of the manuscript (more recent 
detailed discussions can be found in Xiao et al. 2012, Zahid et al. 2013).  
We discuss one simple result about the dust extinction. Based on the 
study of starforming galaxies, the mean dust extinction E(B-V) is 
around 0.4 and smaller than 1 (due to much weak H$\beta$). In other words, 
if our procedure from the subsection 2.1 is available for the spectral 
decomposition, the parameter of the dust extinction should be 
around 0.4 (see results in Xiao et al. 2012, Zahid et al. 2013 and 
references therein). Figure~\ref{ebv} shows the distributions of the dust 
extinctions determined by the procedures with and without the 
considerations of the component $P_{AGN, r_{\lambda}}$. It is clear 
that E(B-V) is around 0.3 with the considerations of 
$P_{AGN, r_{\lambda}}$, but E(B-V) is larger than 1 without the 
considerations of the $P_{AGN, r_{\lambda}}$. Therefore, the AGN component 
$P_{AGN, r_{\lambda}}$ is physically necessary and reliable.

\begin{figure*}
\centering\includegraphics[width = 14cm,height=8cm]{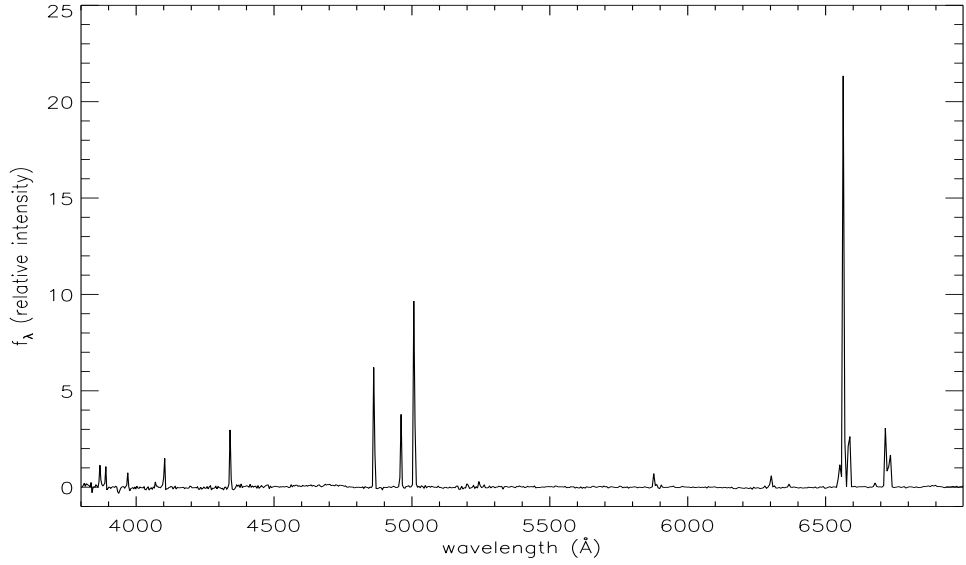}
\caption{The mean spectrum of the 281 reliable candidates for the 
BLR-less AGNs, with AGN continuum components and the stellar 
components having been subtracted.
}
\label{mean}
\end{figure*}

    Last but not least, the final point we should make is that there 
should be no broad emission lines in the SDSS spectra for the 281 
candidates for the BLR-less AGNs, based on the definition of BLR-less AGNs. 
Besides the fitted results as discussed in the subsection 2.1. The final 
mean spectrum of the candidates for the BLR-less AGNs is discussed. 
Here, the mean spectrum is determined by the PCA technique (Principal 
Components Analysis or Karhunen-Loeve Transform method) applied for all 
the 281 SDSS spectra with the power law continuum and the stellar 
components having been subtracted. Then the first principal component 
should represent the mean emission line spectrum of the 281 objects. 
Figure~\ref{mean} shows the mean spectrum for the candidates for the 281 
BLR-less AGNs. It is clear that there are no broad emission lines. In 
other words, we can confirm the selected 281 objects have no broad 
line with the accepted SDSS spectral quality.    

\begin{figure*}
\centering\includegraphics[width = 17cm,height=9cm]{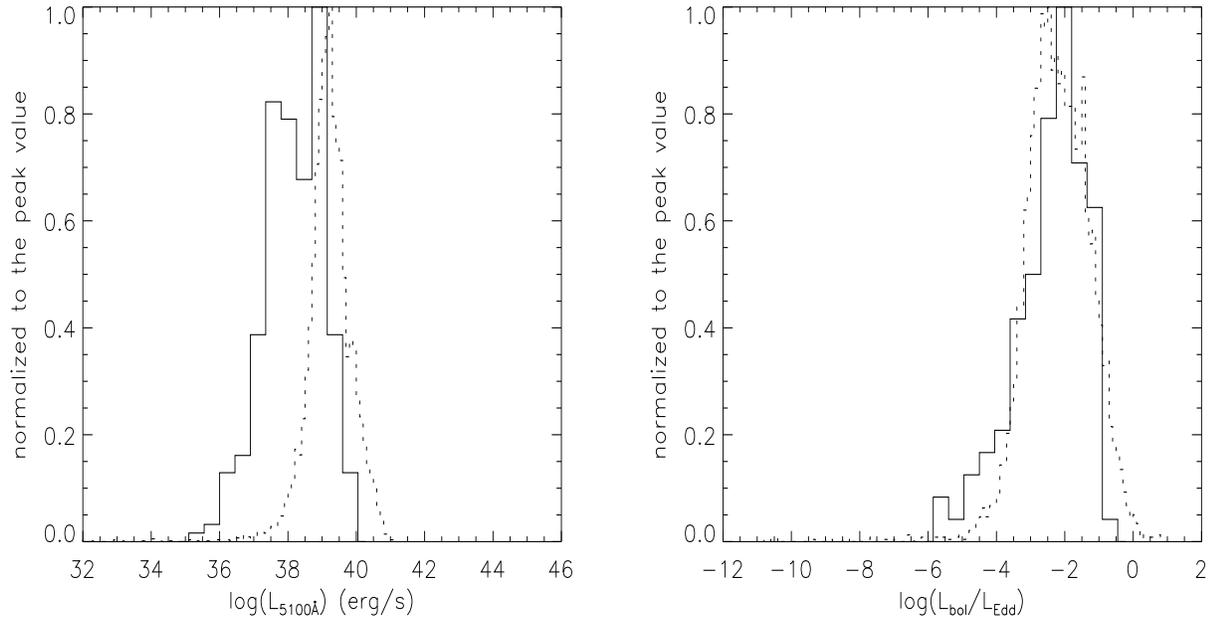}
\caption{The properties of the continuum luminosity and the Eddington 
accretion rate for the reliable candidates for the BLR-less AGNs. 
The solid lines are for the BLR-less AGNs, the dotted lines are 
for the normal broad line AGNs.
}
\label{lum}
\end{figure*}

   Based on the discussions above, we believe although some part of 
the candidates for the BLR-less AGNs have been lost, due to the 
spectral quality, the quality of the SDSS Stripe light curves and the 
strict criteria shown in Equation (5), the selected final candidates 
for the BLR-less AGNs have high confidence levels.  Now, some 
following basic discussions can be shown for the properties the 
candidates for the BLR-less AGNs. We firstly check the dependence of 
the BLR-less AGNs on the luminosity and the accretion rate as discussed 
in the introduction. Figure~\ref{lum} shows the distributions of the 
continuum luminosity at 5100$\AA$ for the candidates for the BLR-less 
AGNs and for the broad line AGNs. The mean values for the continuum 
luminosities at 5100\AA are $10^{39.18\pm0.76} {\rm erg/s}$ and 
$10^{38.18\pm0.85} {\rm erg/s}$ for the broad line AGNs and the candidates 
for the BLR-less AGNs respectively. The luminosity ratio between the 
BLR-less AGNs and normal broad line AGNs is similar as the value shown 
in Tran (2003), but with larger scatter.

    Then, the accretion rate described by  the dimensionless Eddington 
parameter is checked for the normal broad line AGNs and for the candidates 
for the BLR-less AGNs ($\dot{M} =L_{bol}/L_{Eddington}$), and shown in 
Figure~\ref{lum}. The mean values of the dimensionless Eddington accretion 
rate are $10^{-2.19\pm0.98}$ and $10^{-2.44\pm1.02}$ for the broad line AGNs 
and the candidates for the BLR-less AGNs respectively. Here, the 
bolometric luminosity is determined by the AGN continuum luminosity 
(Elvis et al. 1994, Laor 2000, Netzer 2003, Vestergaard 2004, 
Richards et al. 2006, Marconi et al. 2008), and the Eddington luminosity 
is determined by the black hole mass through the M-sigma relation 
(Gebhardt et al., 2000, Ferrarese \& Merritt 2001, Tremaine et al. 2002, 
G\"ultekin et al. 2009, Woo et al. 2010). The result indicates the 
accretion rates for the BLR-less AGNs and the normal AGNs are not much 
different.

\vspace{8mm}

    Finally, our conclusions are as follows. Based on the SSP templates, 
the spectra of all the galaxies and QSOs in the SDSS Stripe82 region 
have been analysed, and then 22693 pure narrow line objects with apparent 
power law continuum components ($R_{AGN}>0.3$ and $R_{ssp} > 0.3$) but 
no broad emission lines are selected to make up of the spectroscopic 
sample for the candidates for the BLR-less AGNs. Then, the properties of t
he photometry variability are checked for the objects in the spectroscopic 
sample. By the properties of the RMS magnitude deviation ($RMS$) and the 
Pearson's coefficients ($R_{1, 2}$) between two different SDSS band 
light curves: $RMS_k>3\times RMS_{M_k}$ and $R_{1, 2}>\sim0.8$, the final 
281 objects are our selected reliable candidates for the BLR-less AGNs, 
which have reliable photometric variabilities and have reliable AGN 
continuum emission but do not have broad emission lines. The reported 
sample four-times enlarges the current sample of the BLR-less AGNs, 
and will provide more reliable information to explain the lack of the 
BLRs of AGNs in our following studies. 

\section*{Acknowledgments}  
Z-XG gratefully acknowledges the anonymous referee for giving us 
constructive comments and suggestions to greatly improve our paper. 
Z-XG acknowledges the kind support from the Chinese grant NSFC-11003043 
and NSFC-11178003. This paper 
has made use of the data from the SDSS projects. Funding for the 
creation and the distribution of the SDSS Archive has been provided 
by the Alfred P. Sloan Foundation, the Participating Institutions, 
the National Aeronautics and Space Administration, the National 
Science Foundation, the U.S. Department of Energy, the Japanese 
Monbukagakusho, and the Max Planck Society. The SDSS is managed by 
the Astrophysical Research Consortium (ARC) for the Participating 
Institutions. The Participating Institutions are The University of 
Chicago, Fermilab, the Institute for Advanced Study, the Japan 
Participation Group, The Johns Hopkins University, Los Alamos 
National Laboratory, the Max-Planck-Institute for Astronomy (MPIA),
the Max-Planck-Institute for Astrophysics (MPA), New
Mexico State University, Princeton University, the United
States Naval Observatory, and the University of Washington.

\clearpage

\begin{table*}
\small
\renewcommand{\tabcolsep}{0.7mm}
\begin{minipage}{175mm}
\caption{Basic Parameters of the 281 candidates for the BLR-less AGNs}
\begin{tabular}{ccccccccl|ccccccccl}
\hline
mpf & z & ra & dec & mag & $R$ & $k$ & $R_{1,2}$ & cl &
mpf & z & ra & dec & mag & $R$ & $k$ & $R_{1,2}$ & cl \\
\hline
51782-0391-236 & 0.080 & 6.91631 & -0.61917 & 17.1 & 0.38 & 9.41 & 0.99 & hii & 
51782-0391-460 & 0.069 & 6.88574 & 0.723382 & 17.3 & 0.31 & 8.30 & 0.98 & non \\
51782-0391-625 & 0.013 & 8.31179 & 0.201018 & 15.6 & 0.47 & 23.7 & 0.99 & hii & 
51783-0385-214 & 0.019 & 355.184 & -0.88746 & 17.8 & 0.58 & 10.4 & 0.98 & hii \\
51783-0395-413 & 0.007 & 14.4857 & 0.869196 & 17.1 & 0.40 & 22.0 & 0.97 & non & 
51788-0373-490 & 0.116 & 331.793 & 0.537971 & 17.7 & 0.42 & 6.33 & 0.76 & non \\
51788-0401-400 & 0.052 & 24.7862 & 0.349467 & 17.3 & 0.48 & 20.6 & 0.99 & non & 
51788-0401-547 & 0.084 & 26.0518 & 0.517921 & 17.6 & 0.38 & 3.28 & 0.89 & non \\
51789-0379-160 & 0.127 & 343.729 & 0.065114 & 18.9 & 0.35 & 7.73 & 0.97 & non & 
51789-0379-537 & 0.025 & 343.643 & 0.830080 & 16.1 & 0.38 & 11.7 & 0.97 & hii \\
51789-0398-127 & 0.073 & 19.6826 & -0.88279 & 20.3 & 0.48 & 23.4 & 0.93 & hii & 
51789-0398-180 & 0.047 & 19.7441 & -0.43511 & 16.9 & 0.32 & 6.09 & 0.96 & hii \\
51789-0398-501 & 0.089 & 19.3185 & 0.502616 & 17.5 & 0.36 & 20.5 & 0.98 & non & 
51791-0374-419 & 0.090 & 333.474 & 0.880517 & 17.3 & 0.54 & 3.79 & 0.92 & non \\
51793-0388-119 & 0.074 & 2.54452 & -0.31131 & 17.6 & 0.48 & 6.92 & 0.94 & non & 
51793-0388-614 & 0.108 & 2.78680 & 0.845443 & 17.1 & 0.53 & 4.79 & 0.89 & hii \\
51793-0392-262 & 0.116 & 8.55825 & 0.039864 & 17.5 & 0.34 & 6.67 & 0.99 & non & 
51793-0392-284 & 0.006 & 8.34204 & -1.12137 & 16.3 & 0.42 & 59.0 & 0.90 & hii \\
51794-0397-008 & 0.018 & 18.8170 & -1.10024 & 16.8 & 0.42 & 6.97 & 0.94 & non & 
51794-0397-336 & 0.003 & 17.2832 & 1.120985 & 16.4 & 0.62 & 7.76 & 0.97 & hii \\
51795-0389-064 & 0.069 & 4.61792 & -0.56956 & 16.6 & 0.42 & 6.05 & 0.94 & hii & 
51811-0381-353 & 0.054 & 346.277 & 0.410307 & 17.9 & 0.37 & 5.08 & 0.89 & hii \\
51812-0394-372 & 0.042 & 12.2676 & 0.765777 & 17.6 & 0.30 & 16.4 & 0.97 & non & 
51812-0394-442 & 0.066 & 12.9475 & 0.989699 & 16.3 & 0.39 & 10.4 & 0.83 & non \\
51812-0404-462 & 0.076 & 31.0442 & 0.526773 & 18.4 & 0.44 & 5.36 & 0.94 & hii & 
51816-0382-010 & 0.024 & 350.466 & -0.69467 & 18.2 & 0.48 & 24.5 & 0.97 & hii \\
51816-0382-083 & 0.091 & 349.658 & -1.22573 & 17.2 & 0.33 & 7.33 & 0.95 & hii & 
51816-0382-185 & 0.024 & 348.901 & -0.45154 & 16.8 & 0.38 & 4.47 & 0.97 & non \\
51816-0382-564 & 0.030 & 349.966 & 1.218043 & 16.3 & 0.47 & 4.25 & 0.95 & non & 
51816-0390-348 & 0.017 & 4.47590 & 0.329340 & 17.0 & 0.31 & 16.3 & 0.99 & hii \\
51816-0390-409 & 0.016 & 5.10742 & 0.826350 & 14.0 & 0.35 & 15.7 & 0.99 & agn & 
51816-0396-011 & 0.076 & 16.9530 & -1.01500 & 18.3 & 0.33 & 12.2 & 0.98 & hii \\
51816-0396-204 & 0.019 & 16.1731 & -0.95594 & 15.0 & 0.39 & 22.3 & 0.99 & hii & 
51816-0396-271 & 0.067 & 15.4541 & -0.15510 & 17.8 & 0.41 & 4.20 & 0.97 & hii \\
51817-0399-196 & 0.043 & 20.7603 & -0.32410 & 15.8 & 0.35 & 5.25 & 0.98 & non & 
51817-0399-323 & 0.008 & 20.5265 & 0.938026 & 19.8 & 0.38 & 26.3 & 0.96 & hii \\
51817-0399-324 & 0.007 & 20.5577 & 0.958729 & 17.3 & 0.51 & 48.7 & 0.96 & hii & 
51817-0399-434 & 0.121 & 20.7876 & 0.166059 & 18.2 & 0.65 & 16.0 & 0.97 & agn \\
51817-0399-469 & 0.067 & 20.8725 & 0.568883 & 16.5 & 0.33 & 4.31 & 0.97 & hii & 
51817-0406-066 & 0.021 & 36.5963 & -0.49192 & 17.4 & 0.44 & 12.8 & 0.95 & non \\
51817-0411-119 & 0.029 & 46.5573 & -0.34154 & 16.4 & 0.37 & 19.7 & 0.99 & non & 
51818-0383-222 & 0.115 & 350.824 & -0.23943 & 16.8 & 0.51 & 9.02 & 0.98 & non \\
51820-0407-023 & 0.245 & 39.1133 & -0.61284 & 18.8 & 0.33 & 10.9 & 0.94 & non & 
51820-0407-320 & 0.021 & 36.5963 & -0.49192 & 17.4 & 0.35 & 12.8 & 0.95 & hii \\
51820-0407-509 & 0.021 & 38.1587 & 0.594263 & 15.6 & 0.63 & 5.02 & 0.97 & hii & 
51821-0384-257 & 0.083 & 352.805 & -0.93731 & 17.7 & 0.50 & 28.7 & 0.90 & hii \\
51821-0384-278 & 0.008 & 352.758 & -0.13197 & 17.2 & 0.40 & 23.4 & 0.99 & non & 
51869-0406-063 & 0.021 & 36.5963 & -0.49192 & 17.4 & 0.36 & 12.8 & 0.95 & hii \\
51871-0403-237 & 0.027 & 29.0007 & -0.36291 & 17.0 & 0.32 & 20.5 & 0.98 & hii & 
51871-0403-452 & 0.081 & 29.5011 & 1.089686 & 17.1 & 0.32 & 4.16 & 0.96 & hii \\
51871-0403-621 & 0.111 & 31.0741 & 0.710236 & 17.2 & 0.34 & 8.88 & 0.92 & hii & 
51871-0409-114 & 0.021 & 42.4203 & -0.52336 & 14.5 & 0.32 & 9.19 & 0.95 & hii \\
51871-0409-237 & 0.005 & 41.6048 & -0.49868 & 12.2 & 0.46 & 45.2 & 0.99 & hii & 
51871-0409-298 & 0.043 & 41.2430 & -0.94640 & 16.7 & 0.31 & 23.0 & 0.90 & hii \\
51871-0409-465 & 0.075 & 41.9021 & 0.451147 & 17.8 & 0.48 & 3.52 & 0.89 & non & 
51871-0412-075 & 0.025 & 48.4493 & -0.24329 & 18.2 & 0.67 & 40.9 & 0.99 & non \\
51871-0412-581 & 0.025 & 48.3041 & 0.282182 & 17.5 & 0.34 & 11.4 & 0.92 & hii & 
51873-0411-101 & 0.029 & 46.5573 & -0.34154 & 16.4 & 0.36 & 19.7 & 0.99 & hii \\
51876-0394-489 & 0.115 & 13.0620 & 0.583035 & 16.9 & 0.40 & 20.7 & 0.94 & agn & 
51876-0394-508 & 0.147 & 13.5640 & 0.271241 & 18.3 & 0.31 & 4.70 & 0.92 & non \\
51876-0394-565 & 0.041 & 13.5639 & 1.077725 & 16.1 & 0.36 & 18.8 & 0.99 & non & 
51876-0406-063 & 0.021 & 36.5963 & -0.49192 & 17.4 & 0.43 & 12.8 & 0.95 & non \\
51877-0385-211 & 0.019 & 355.184 & -0.88746 & 17.8 & 0.42 & 10.4 & 0.98 & hii & 
51877-0404-470 & 0.076 & 31.0442 & 0.526773 & 18.4 & 0.60 & 5.36 & 0.94 & hii \\
51900-0390-350 & 0.017 & 4.47590 & 0.329340 & 17.0 & 0.34 & 16.3 & 0.99 & hii & 
51900-0390-407 & 0.016 & 5.10742 & 0.826350 & 14.0 & 0.32 & 15.7 & 0.99 & agn \\
51900-0406-063 & 0.021 & 36.5963 & -0.49192 & 17.4 & 0.47 & 12.8 & 0.95 & hii & 
51913-0394-569 & 0.041 & 13.5639 & 1.077725 & 16.1 & 0.40 & 18.8 & 0.99 & hii \\
51929-0413-196 & 0.066 & 49.4594 & -0.12439 & 16.1 & 0.32 & 3.05 & 0.94 & hii & 
51929-0413-356 & 0.025 & 48.3041 & 0.282182 & 17.5 & 0.45 & 11.4 & 0.92 & hii \\
51931-0412-009 & 0.027 & 48.6088 & -1.14626 & 17.3 & 0.33 & 7.95 & 0.97 & non & 
51931-0412-588 & 0.025 & 48.3041 & 0.282182 & 17.5 & 0.39 & 11.4 & 0.92 & hii \\
51936-0412-009 & 0.027 & 48.6088 & -1.14626 & 17.3 & 0.52 & 7.95 & 0.97 & non & 
51936-0412-582 & 0.025 & 48.3041 & 0.282182 & 17.5 & 0.44 & 11.4 & 0.92 & hii \\
51942-0412-016 & 0.027 & 48.6088 & -1.14626 & 17.3 & 0.33 & 7.95 & 0.97 & hii & 
51942-0412-582 & 0.025 & 48.3041 & 0.282182 & 17.5 & 0.56 & 11.4 & 0.92 & hii \\
52078-0371-243 & 0.122 & 327.225 & -1.16658 & 18.9 & 0.31 & 4.27 & 0.89 & hii & 
52143-0376-576 & 0.055 & 338.028 & 0.915153 & 17.8 & 0.53 & 6.63 & 0.96 & non \\
52145-0377-372 & 0.059 & 338.816 & 0.892289 & 17.2 & 0.31 & 3.45 & 0.84 & hii & 
52146-0378-104 & 0.015 & 341.767 & -0.07408 & 14.5 & 0.34 & 31.7 & 0.98 & non \\
52146-0378-277 & 0.058 & 340.269 & -0.07247 & 17.8 & 0.34 & 22.6 & 0.99 & hii & 
52146-0378-347 & 0.005 & 340.391 & 0.400612 & 15.1 & 0.38 & 55.2 & 0.99 & non \\
52174-0676-265 & 0.212 & 342.081 & -0.61156 & 17.9 & 0.31 & 5.24 & 0.95 & non & 
52174-0676-408 & 0.048 & 343.019 & 1.246003 & 18.1 & 0.40 & 19.5 & 0.93 & hii \\
52175-0708-264 & 0.021 & 42.4203 & -0.52336 & 14.5 & 0.30 & 9.19 & 0.95 & hii & 
52175-0708-419 & 0.059 & 43.2783 & 1.161945 & 17.3 & 0.36 & 11.1 & 0.96 & hii \\
52177-0679-157 & 0.233 & 348.889 & -0.10358 & 19.9 & 0.38 & 3.77 & 0.93 & hii & 
52177-0707-507 & 0.075 & 41.9021 & 0.451147 & 17.8 & 0.48 & 3.52 & 0.89 & hii \\
52178-0676-360 & 0.085 & 341.789 & 0.175907 & 17.0 & 0.32 & 5.74 & 0.98 & non & 
52178-0676-410 & 0.048 & 343.019 & 1.246003 & 18.1 & 0.60 & 19.5 & 0.93 & hii \\
52178-0676-532 & 0.025 & 343.643 & 0.830053 & 16.1 & 0.38 & 11.7 & 0.97 & hii & 
52178-0676-552 & 0.127 & 343.729 & 0.065114 & 18.9 & 0.40 & 7.73 & 0.97 & hii \\
52199-0681-139 & 0.023 & 352.971 & -0.82643 & 16.9 & 0.39 & 56.8 & 0.99 & hii & 
52199-0691-527 & 0.064 & 11.7256 & 1.011370 & 17.4 & 0.32 & 8.03 & 0.99 & non \\
52200-0680-166 & 0.115 & 350.824 & -0.23943 & 16.8 & 0.35 & 9.02 & 0.98 & agn & 
52200-0705-486 & 0.326 & 37.6558 & 0.189943 & 19.9 & 0.30 & 6.25 & 0.92 & non \\
52201-0674-347 & 0.059 & 337.764 & 0.283131 & 17.7 & 0.33 & 24.8 & 0.99 & hii & 
52201-0692-278 & 0.309 & 12.6371 & -0.39110 & 19.4 & 0.46 & 5.34 & 0.93 & hii \\
52202-0695-105 & 0.047 & 19.6483 & -0.23312 & 15.1 & 0.33 & 22.8 & 0.99 & hii & 
52202-0695-325 & 0.003 & 18.2318 & 0.991891 & 18.0 & 0.66 & 27.5 & 0.97 & hii \\
52202-0695-413 & 0.064 & 18.5338 & 0.846754 & 17.7 & 0.37 & 4.68 & 0.86 & hii & 
52202-0699-432 & 0.084 & 26.0518 & 0.517921 & 17.6 & 0.33 & 3.28 & 0.89 & hii \\
52202-0711-485 & 0.022 & 49.3344 & -0.07717 & 16.6 & 0.34 & 60.2 & 0.99 & hii & 
52203-0688-304 & 0.069 & 4.61792 & -0.56956 & 16.6 & 0.36 & 6.05 & 0.94 & agn \\
52203-0688-391 & 0.105 & 5.12084 & 0.300712 & 18.6 & 0.35 & 3.09 & 0.91 & hii & 
52205-0704-383 & 0.291 & 34.9974 & 0.159337 & 19.1 & 0.45 & 8.56 & 0.78 & non \\
52205-0704-482 & 0.327 & 35.9311 & 0.552009 & 19.5 & 0.50 & 13.3 & 0.99 & hii & 
52205-0709-065 & 0.119 & 46.5204 & -0.46675 & 18.1 & 0.34 & 9.79 & 0.97 & non \\
52209-0696-406 & 0.055 & 20.5731 & 1.007560 & 16.1 & 0.44 & 11.1 & 0.96 & hii & 
52209-0696-515 & 0.056 & 21.1558 & 0.087213 & 17.9 & 0.32 & 21.9 & 0.88 & hii \\
52226-0697-508 & 0.080 & 22.8879 & 0.552905 & 19.6 & 0.39 & 24.6 & 0.94 & non & 
52235-0412-079 & 0.025 & 48.4493 & -0.24329 & 18.2 & 0.62 & 40.9 & 0.99 & non \\
52235-0412-587 & 0.025 & 48.3041 & 0.282182 & 17.5 & 0.36 & 11.4 & 0.92 & hii & 
52250-0412-586 & 0.025 & 48.3041 & 0.282182 & 17.5 & 0.33 & 11.4 & 0.92 & hii \\
52254-0412-004 & 0.027 & 48.6088 & -1.14626 & 17.3 & 0.44 & 7.95 & 0.97 & hii & 
52254-0412-072 & 0.025 & 48.4493 & -0.24329 & 18.2 & 0.56 & 40.9 & 0.99 & hii \\
\hline
\end{tabular}\\
Notice: The first and the tenth columns list the SDSS MJD-PLATE-FIBERID, 
the second and eleventh columns show the redshift, the third and twelfth 
columns show the ra value in degree, the firth and thirteenth columns 
show the dec values in degree, the sixth and fourteenth columns show 
the magnitudes in r band, the seventh and fifteenth columns show the 
parameter $R_{AGN}$, the eighth and sixteenth columns show the maximum 
values of $k_{RMS}=RMS_{k}/RMS_{M_k}$ in the SDSS gri bands, the ninth 
and seventeenth columns show maximum values of Spearman's correlation 
coefficient $R_{1,2}$, and the tenth and final columns show the 
classifications for the objects: AGNs (agn), HII galaxies 
(hii) or non-classified objects (non).
\end{minipage}
\end{table*}

\setcounter{table}{0}
\begin{table*}
\small
\renewcommand{\tabcolsep}{0.7mm}
\begin{minipage}{175mm}
\caption{--continued.}
\begin{tabular}{ccccccccl|ccccccccl}
\hline
mpf & z & ra & dec & mag & $R$ & $k$ & $R_{1,2}$ & cl &
mpf & z & ra & dec & mag & $R$ & $k$ & $R_{1,2}$ & cl \\
\hline
52254-0412-591 & 0.025 & 48.3041 & 0.282182 & 17.5 & 0.36 & 11.4 & 0.92 & hii & 
52254-0693-367 & 0.018 & 14.7013 & 0.589705 & 18.5 & 0.45 & 57.4 & 0.99 & non \\
52258-0412-079 & 0.025 & 48.4493 & -0.24329 & 18.2 & 0.68 & 40.9 & 0.99 & hii & 
52258-0412-587 & 0.025 & 48.3041 & 0.282182 & 17.5 & 0.37 & 11.4 & 0.92 & hii \\
52261-0690-587 & 0.218 & 10.7037 & 0.578405 & 18.8 & 0.43 & 4.78 & 0.94 & hii & 
52261-0690-609 & 0.018 & 10.4392 & 1.174548 & 17.9 & 0.45 & 37.3 & 0.98 & non \\
52262-0689-064 & 0.067 & 8.37281 & -0.54093 & 18.0 & 0.48 & 11.0 & 0.97 & non & 
52262-0689-070 & 0.106 & 8.37799 & -0.32262 & 18.8 & 0.54 & 4.35 & 0.92 & hii \\
52264-0803-154 & 0.138 & 47.8015 & -0.28485 & 18.5 & 0.36 & 10.3 & 0.96 & hii & 
52264-0803-320 & 0.119 & 46.5204 & -0.46675 & 18.1 & 0.40 & 9.79 & 0.97 & non \\
52286-0804-294 & 0.141 & 48.7204 & -0.64079 & 18.6 & 0.31 & 9.03 & 0.92 & non & 
52286-0804-429 & 0.151 & 49.0715 & 0.465250 & 18.9 & 0.47 & 7.36 & 0.89 & non \\
52289-0802-348 & 0.043 & 44.8906 & 0.915088 & 18.6 & 0.51 & 10.8 & 0.89 & non & 
52289-0802-453 & 0.199 & 45.3461 & 0.476043 & 18.9 & 0.47 & 6.06 & 0.89 & hii \\
52297-0803-561 & 0.209 & 48.3019 & 0.532733 & 18.1 & 0.38 & 3.92 & 0.91 & non & 
52297-0803-600 & 0.156 & 48.2579 & 0.725732 & 20.9 & 0.44 & 4.11 & 0.85 & non \\
52318-0803-136 & 0.138 & 47.8015 & -0.28485 & 18.5 & 0.30 & 10.3 & 0.96 & hii & 
52318-0803-290 & 0.119 & 46.5204 & -0.46675 & 18.1 & 0.39 & 9.79 & 0.97 & hii \\
52318-0803-334 & 0.138 & 46.3641 & 0.128511 & 19.3 & 0.51 & 4.96 & 0.84 & non & 
52435-0981-046 & 0.080 & 310.852 & -0.17299 & 17.1 & 0.34 & 12.6 & 0.98 & hii \\
52443-0983-360 & 0.130 & 312.062 & 0.145880 & 17.8 & 0.44 & 6.60 & 0.95 & non & 
52465-0990-307 & 0.063 & 324.507 & -0.40264 & 17.4 & 0.32 & 14.3 & 0.99 & non \\
52465-0990-443 & 0.068 & 325.760 & 1.242723 & 17.1 & 0.39 & 3.16 & 0.86 & non & 
52466-0982-127 & 0.118 & 312.025 & -0.23095 & 17.8 & 0.37 & 13.5 & 0.99 & non \\
52468-0989-230 & 0.120 & 323.220 & -0.39948 & 17.4 & 0.34 & 13.5 & 0.99 & hii & 
52518-0687-534 & 0.303 & 4.27962 & 0.966183 & 19.4 & 0.55 & 5.32 & 0.91 & hii \\
52520-0670-308 & 0.300 & 15.2634 & -0.36469 & 19.8 & 0.38 & 5.03 & 0.84 & non & 
52520-0988-498 & 0.063 & 322.660 & 0.254061 & 17.4 & 0.44 & 9.73 & 0.89 & hii \\
52520-0988-560 & 0.049 & 322.772 & 0.953709 & 17.8 & 0.33 & 5.40 & 0.96 & hii & 
52521-1095-198 & 0.008 & 351.088 & -0.10817 & 18.6 & 0.63 & 7.88 & 0.94 & hii \\
52521-1095-349 & 0.031 & 349.876 & 0.371450 & 19.2 & 0.58 & 38.4 & 0.98 & hii & 
52523-0684-042 & 0.111 & 358.723 & -1.08919 & 19.2 & 0.40 & 5.10 & 0.94 & hii \\
52523-1082-567 & 0.047 & 16.8977 & 1.194968 & 18.0 & 0.42 & 3.42 & 0.94 & hii & 
52524-1022-156 & 0.120 & 312.239 & -0.43634 & 18.5 & 0.48 & 11.9 & 0.90 & hii \\
52524-1022-227 & 0.216 & 311.452 & -0.60317 & 18.3 & 0.30 & 6.01 & 0.90 & non & 
52524-1022-415 & 0.027 & 311.455 & 0.674695 & 17.7 & 0.46 & 6.12 & 0.80 & hii \\
52525-1034-601 & 0.112 & 334.940 & 0.375554 & 18.9 & 0.42 & 5.28 & 0.95 & non & 
52525-1034-625 & 0.057 & 334.453 & 1.070917 & 19.5 & 0.37 & 5.96 & 0.96 & non \\
52527-0669-307 & 0.283 & 0.66073 & -0.20296 & 19.6 & 0.33 & 5.58 & 0.87 & hii & 
52531-1081-048 & 0.005 & 18.8768 & -0.86274 & 16.5 & 0.49 & 70.6 & 0.93 & non \\
52531-1085-368 & 0.017 & 10.4457 & 1.172180 & 18.7 & 0.53 & 40.0 & 0.98 & hii & 
52557-1027-034 & 0.293 & 321.683 & -0.66242 & 19.5 & 0.61 & 3.55 & 0.82 & non \\
52557-1027-111 & 0.124 & 320.953 & -0.91073 & 18.5 & 0.32 & 11.6 & 0.96 & non & 
52557-1027-200 & 0.124 & 320.899 & -0.89219 & 19.0 & 0.49 & 5.06 & 0.89 & hii \\
52557-1027-309 & 0.010 & 319.555 & -0.86333 & 18.3 & 0.33 & 6.14 & 0.90 & non & 
52557-1027-452 & 0.086 & 320.542 & 0.392080 & 18.5 & 0.52 & 19.3 & 0.98 & non \\
52558-1026-178 & 0.340 & 318.844 & -0.20190 & 19.0 & 0.31 & 5.91 & 0.93 & non & 
52558-1026-280 & 0.060 & 317.996 & -0.22021 & 17.2 & 0.39 & 7.69 & 1.00 & agn \\
52559-0669-301 & 0.283 & 0.66073 & -0.20296 & 19.6 & 0.61 & 5.58 & 0.87 & non & 
52562-1028-489 & 0.030 & 322.438 & 0.678383 & 19.4 & 0.59 & 12.3 & 0.98 & hii \\
52582-1036-211 & 0.059 & 337.244 & -1.00362 & 18.4 & 0.38 & 10.7 & 0.98 & hii & 
52582-1036-271 & 0.055 & 336.923 & -0.15137 & 18.1 & 0.39 & 3.87 & 0.93 & hii \\
52582-1036-528 & 0.037 & 337.988 & 1.095786 & 17.7 & 0.45 & 18.7 & 0.98 & hii & 
52589-1066-160 & 0.138 & 47.8020 & -0.28447 & 20.9 & 0.33 & 3.75 & 0.75 & non \\
52590-0675-225 & 0.127 & 340.831 & -0.35270 & 18.0 & 0.36 & 5.34 & 0.96 & hii & 
52590-0675-331 & 0.089 & 340.445 & 1.201112 & 17.5 & 0.41 & 4.54 & 0.99 & hii \\
52590-0675-540 & 0.016 & 341.517 & 0.937274 & 17.5 & 0.52 & 11.3 & 0.98 & non & 
52590-1069-193 & 0.005 & 41.6056 & -0.49924 & 17.2 & 0.48 & 30.9 & 0.96 & hii \\
52590-1069-621 & 0.028 & 42.9714 & 0.665730 & 20.7 & 0.49 & 10.0 & 0.92 & hii & 
52591-1070-536 & 0.023 & 40.2009 & 1.101998 & 19.9 & 0.69 & 16.4 & 0.85 & hii \\
52591-1084-099 & 0.062 & 12.8838 & -0.76888 & 19.0 & 0.62 & 10.4 & 0.82 & hii & 
52591-1084-407 & 0.015 & 12.4672 & 0.963360 & 14.6 & 0.34 & 3.47 & 0.96 & hii \\
52616-1067-066 & 0.024 & 46.3729 & -0.38143 & 16.8 & 0.43 & 70.4 & 0.85 & hii & 
52637-1179-039 & 0.027 & 48.6088 & -1.14626 & 17.3 & 0.34 & 7.95 & 0.97 & hii \\
52637-1179-048 & 0.025 & 48.4493 & -0.24329 & 18.2 & 0.49 & 40.9 & 0.99 & hii & 
52639-1092-600 & 0.081 & 358.352 & 0.060815 & 18.4 & 0.41 & 14.0 & 0.99 & hii \\
52641-1071-457 & 0.021 & 37.6254 & 0.379881 & 18.1 & 0.39 & 10.9 & 0.96 & hii & 
52643-1078-118 & 0.015 & 23.8070 & -1.17728 & 18.6 & 0.41 & 5.33 & 0.87 & hii \\
52643-1078-458 & 0.016 & 23.2780 & 0.177129 & 19.7 & 0.68 & 20.3 & 0.90 & hii & 
52669-0811-252 & 0.037 & 47.1443 & -0.85147 & 19.3 & 0.48 & 7.45 & 0.91 & non \\
52669-0811-415 & 0.030 & 47.2661 & 0.646299 & 19.3 & 0.48 & 6.69 & 0.92 & hii & 
52669-0811-490 & 0.048 & 47.8897 & 0.533254 & 18.4 & 0.53 & 9.67 & 0.98 & hii \\
52669-0811-634 & 0.151 & 49.0715 & 0.465250 & 18.9 & 0.46 & 7.36 & 0.89 & non & 
52797-1023-295 & 0.129 & 312.590 & -1.00562 & 18.6 & 0.37 & 7.31 & 0.97 & non \\
52797-1023-389 & 0.200 & 312.525 & 0.558429 & 18.8 & 0.36 & 15.7 & 0.75 & agn & 
52797-1023-510 & 0.201 & 313.706 & 0.697158 & 19.5 & 0.34 & 4.68 & 0.81 & non \\
52813-1034-254 & 0.131 & 333.099 & -0.96498 & 18.1 & 0.32 & 6.24 & 0.99 & non & 
52813-1034-604 & 0.057 & 334.453 & 1.070917 & 19.5 & 0.31 & 5.96 & 0.96 & non \\
52816-1035-031 & 0.090 & 336.789 & -0.62212 & 18.1 & 0.40 & 16.6 & 0.99 & hii & 
52816-1035-572 & 0.057 & 336.033 & 0.861933 & 18.2 & 0.37 & 13.0 & 0.99 & hii \\
52822-1033-140 & 0.095 & 331.749 & -0.88593 & 18.9 & 0.31 & 8.36 & 0.99 & hii & 
52822-1033-198 & 0.207 & 331.462 & -0.11459 & 19.6 & 0.58 & 4.68 & 0.90 & hii \\
52826-1024-216 & 0.133 & 314.818 & -0.78427 & 19.1 & 0.39 & 6.30 & 0.90 & non & 
52826-1024-365 & 0.027 & 314.454 & 1.079220 & 18.9 & 0.47 & 7.98 & 0.92 & hii \\
52826-1024-536 & 0.056 & 315.207 & 1.190049 & 17.8 & 0.37 & 11.3 & 0.92 & non & 
52826-1037-503 & 0.160 & 339.814 & 0.317021 & 19.6 & 0.48 & 6.48 & 0.94 & hii \\
52826-1037-587 & 0.058 & 340.770 & 0.356819 & 18.1 & 0.31 & 31.6 & 0.75 & non & 
52878-1037-376 & 0.060 & 338.955 & 0.537190 & 19.1 & 0.31 & 22.1 & 0.75 & non \\
52878-1037-604 & 0.059 & 340.390 & 0.817578 & 18.8 & 0.43 & 4.78 & 0.84 & hii & 
52883-1496-181 & 0.059 & 12.4817 & -0.20429 & 18.9 & 0.31 & 9.20 & 0.98 & hii \\
52884-1028-111 & 0.129 & 322.510 & -0.99479 & 18.9 & 0.47 & 14.2 & 0.98 & non & 
52884-1028-391 & 0.031 & 321.707 & 0.975860 & 18.4 & 0.47 & 8.30 & 0.94 & non \\
52884-1028-471 & 0.030 & 322.438 & 0.678383 & 19.4 & 0.47 & 12.3 & 0.98 & hii & 
52886-1497-450 & 0.017 & 14.7330 & 1.005642 & 18.5 & 0.51 & 29.2 & 0.97 & hii \\
52886-1497-529 & 0.017 & 14.9293 & 0.921629 & 19.2 & 0.50 & 17.8 & 0.94 & hii & 
52903-1090-042 & 0.031 & 2.52555 & -0.43833 & 17.8 & 0.51 & 27.8 & 0.96 & non \\
52903-1090-346 & 0.076 & 0.81866 & 1.085157 & 18.3 & 0.44 & 6.65 & 0.79 & hii & 
52903-1475-374 & 0.009 & 331.180 & 1.039517 & 18.3 & 0.59 & 8.56 & 0.88 & non \\
52903-1475-555 & 0.154 & 332.219 & 0.285299 & 20.9 & 0.34 & 4.95 & 0.85 & agn & 
52912-1029-094 & 0.118 & 324.504 & -0.62392 & 18.8 & 0.34 & 5.23 & 0.94 & hii \\
52912-1029-194 & 0.026 & 323.662 & -0.16645 & 18.3 & 0.48 & 6.97 & 0.93 & hii & 
52912-1029-550 & 0.166 & 324.588 & 0.284699 & 19.6 & 0.39 & 3.46 & 0.86 & hii \\
52912-1106-456 & 0.010 & 329.737 & 1.018632 & 18.7 & 0.55 & 19.3 & 0.89 & hii & 
52914-1030-159 & 0.075 & 326.072 & -0.24196 & 19.4 & 0.32 & 7.15 & 0.88 & hii \\
52914-1498-075 & 0.226 & 17.1398 & -0.46871 & 18.8 & 0.49 & 8.59 & 0.94 & hii & 
52930-1087-005 & 0.067 & 7.65952 & -0.92077 & 19.1 & 0.56 & 28.7 & 0.99 & hii \\
52930-1087-112 & 0.040 & 7.41983 & -0.41391 & 18.3 & 0.49 & 7.26 & 0.91 & non & 
52930-1087-123 & 0.066 & 6.87844 & -1.19850 & 18.2 & 0.45 & 5.85 & 0.76 & hii \\
52930-1087-157 & 0.068 & 7.32642 & -0.08815 & 19.7 & 0.48 & 26.3 & 0.99 & hii & 
52930-1087-543 & 0.059 & 7.43631 & 0.776922 & 17.9 & 0.31 & 12.0 & 0.99 & hii \\
52931-1514-187 & 0.227 & 47.4285 & -0.26699 & 19.3 & 0.31 & 4.91 & 0.93 & hii & 
52931-1514-219 & 0.037 & 47.4941 & -0.69401 & 21.2 & 0.47 & 8.77 & 0.87 & hii \\
52931-1514-625 & 0.131 & 49.2046 & 0.084835 & 21.0 & 0.67 & 7.22 & 0.91 & non & 
52932-1515-301 & 0.025 & 48.4502 & -0.24319 & 18.9 & 0.59 & 47.1 & 0.99 & non \\
52932-1515-359 & 0.025 & 48.3037 & 0.282169 & 17.5 & 0.32 & 11.4 & 0.92 & hii & 
52932-1522-500 & 0.183 & 320.470 & 0.468425 & 18.2 & 0.37 & 10.3 & 0.95 & non \\
52933-1075-095 & 0.052 & 29.8693 & -0.96168 & 18.3 & 0.35 & 8.11 & 0.98 & non & 
52933-1075-464 & 0.022 & 29.6694 & 0.522198 & 19.6 & 0.53 & 22.1 & 0.98 & hii \\
52933-1075-546 & 0.040 & 30.3278 & 0.451110 & 17.8 & 0.40 & 5.75 & 0.95 & hii & 
52933-1474-189 & 0.016 & 329.601 & -0.73713 & 20.4 & 0.57 & 10.8 & 0.86 & hii \\
52933-1493-628 & 0.072 & 8.19787 & 0.544704 & 18.7 & 0.40 & 7.40 & 0.94 & hii & 
52937-1494-342 & 0.094 & 7.84317 & 0.309932 & 19.3 & 0.52 & 8.33 & 0.96 & non \\
52937-1494-563 & 0.068 & 9.50234 & 1.120122 & 18.3 & 0.30 & 6.45 & 0.96 & hii & 
52937-1523-262 & 0.070 & 317.716 & -0.43571 & 19.2 & 0.38 & 9.77 & 0.95 & hii \\
52944-1495-402 & 0.079 & 10.3891 & 1.235386 & 0.00 & 0.59 & 12.2 & 0.99 & non & 
52944-1508-492 & 0.163 & 35.8176 & 0.648005 & 18.7 & 0.36 & 6.68 & 0.91 & hii \\
\hline
\end{tabular}\\
\end{minipage}
\end{table*}

\setcounter{table}{0}
\begin{table*}
\small
\renewcommand{\tabcolsep}{0.7mm}
\begin{minipage}{175mm}
\caption{--continued.}
\begin{tabular}{ccccccccl|ccccccccl}
\hline
mpf & z & ra & dec & mag & $R$ & $k$ & $R_{1,2}$ & cl &
mpf & z & ra & dec & mag & $R$ & $k$ & $R_{1,2}$ & cl \\
\hline
52946-1511-290 & 0.054 & 41.2201 & -0.85115 & 18.8 & 0.40 & 5.15 & 0.94 & non & 
52964-1476-602 & 0.057 & 334.453 & 1.070917 & 19.5 & 0.34 & 5.96 & 0.96 & non \\
52964-1487-229 & 0.068 & 355.029 & -0.48348 & 18.8 & 0.39 & 9.85 & 0.95 & hii & 
52965-1488-347 & 0.023 & 356.157 & 0.177964 & 19.0 & 0.44 & 4.61 & 0.86 & hii \\
52974-1096-212 & 0.068 & 348.910 & -0.68242 & 18.8 & 0.31 & 8.58 & 0.86 & non & 
52974-1096-283 & 0.012 & 348.466 & -1.17533 & 18.5 & 0.36 & 7.61 & 0.86 & hii \\
52991-1489-096 & 0.026 & 359.951 & -0.76895 & 19.0 & 0.52 & 8.37 & 0.77 & hii & 
52991-1564-218 & 0.155 & 47.6880 & -0.64722 & 19.1 & 0.32 & 4.57 & 0.95 & hii \\
52993-1486-448 & 0.027 & 353.154 & 1.223620 & 18.8 & 0.53 & 5.72 & 0.87 & non & 
52994-1490-111 & 0.074 & 2.54401 & -0.31029 & 18.8 & 0.44 & 5.44 & 0.98 & hii \\
52996-1491-501 & 0.040 & 3.57930 & 0.200984 & 19.2 & 0.50 & 14.9 & 0.95 & hii & 
53001-1499-014 & 0.005 & 18.8829 & -0.86236 & 16.5 & 0.52 & 70.6 & 0.93 & hii \\
53001-1499-452 & 0.004 & 18.2597 & 0.972991 & 17.0 & 0.38 & 6.02 & 0.87 & hii & 
53001-1499-525 & 0.003 & 18.4185 & 0.877542 & 20.2 & 0.41 & 8.74 & 0.76 & non \\
53001-1499-527 & 0.003 & 18.5847 & 0.916690 & 18.8 & 0.36 & 63.3 & 0.99 & non & 
53001-1499-577 & 0.065 & 18.7107 & 0.803241 & 19.0 & 0.31 & 11.5 & 0.94 & hii \\
53001-1499-596 & 0.043 & 18.8711 & 0.703726 & 18.8 & 0.42 & 21.8 & 0.95 & hii & 
53052-1562-056 & 0.037 & 44.6053 & -1.07781 & 19.1 & 0.49 & 7.19 & 0.89 & hii \\
53172-1031-596 & 0.204 & 328.742 & 0.574362 & 19.6 & 0.42 & 6.37 & 0.94 & non & 
53175-1032-536 & 0.204 & 329.905 & 0.997784 & 19.5 & 0.39 & 5.94 & 0.86 & non \\
53239-1025-120 & 0.050 & 317.415 & -0.35844 & 18.5 & 0.35 & 15.2 & 0.99 & hii & 
53239-1025-513 & 0.050 & 316.929 & 0.343028 & 17.5 & 0.44 & 10.1 & 0.96 & non \\
53239-1025-531 & 0.187 & 317.069 & 0.855431 & 19.4 & 0.44 & 4.35 & 0.78 & hii & 
53271-1558-221 & 0.042 & 35.1198 & -0.43119 & 18.7 & 0.35 & 20.2 & 0.99 & non \\
53271-1558-487 & 0.160 & 35.8396 & 0.570051 & 19.2 & 0.43 & 11.7 & 0.99 & hii & 
53287-1555-270 & 0.059 & 29.0265 & -0.00753 & 18.5 & 0.30 & 6.66 & 0.97 & hii \\
53287-1555-276 & 0.358 & 28.8100 & 0.104836 & 19.7 & 0.32 & 12.4 & 0.98 & non & 
53317-1560-056 & 0.176 & 40.3656 & -0.75473 & 18.7 & 0.30 & 4.65 & 0.80 & non \\
53730-1500-120 & 0.088 & 20.2711 & -0.21799 & 19.2 & 0.41 & 11.0 & 0.76 & hii & 
53730-1500-245 & 0.108 & 19.4387 & -1.17558 & 19.7 & 0.35 & 5.52 & 0.88 & hii \\
53730-1500-579 & 0.007 & 20.5374 & 0.948195 & 15.7 & 0.62 & 86.5 & 0.99 & hii & 
53734-1542-472 & 0.106 & 5.11576 & 0.300462 & 18.6 & 0.44 & 3.09 & 0.91 & hii \\
53740-1501-360 & 0.007 & 20.5515 & 0.948310 & 18.1 & 0.61 & 49.1 & 0.98 & hii & 
53740-1501-412 & 0.016 & 21.4156 & 0.127994 & 19.4 & 0.42 & 26.1 & 0.95 & hii \\
53740-1501-455 & 0.006 & 21.5197 & 0.315263 & 17.4 & 0.60 & 67.4 & 0.99 & non & 
53740-1556-123 & 0.075 & 31.7642 & -1.13393 & 18.8 & 0.50 & 3.16 & 0.86 & non \\
53740-1556-188 & 0.042 & 31.4155 & -0.68968 & 20.1 & 0.51 & 28.2 & 0.98 & hii & 
53741-1502-510 & 0.026 & 23.5799 & 1.236970 & 19.1 & 0.43 & 20.2 & 0.93 & hii \\
53742-1512-479 & 0.041 & 43.7070 & 0.039400 & 20.7 & 0.32 & 14.9 & 0.92 & hii & 
  &   &   &   &   &   &   &   &   \\
\hline
\end{tabular}\\
\end{minipage}
\end{table*}

\clearpage

\label{lastpage}
\end{document}